# Circular phonons generate negligible magnetization in insulators

C-C. Wu[1], M. Buzzi[1], A. Iudica[1], R. Romero III[1], G. De Vecchi[1], R. Merlin[1,2], A. Cavalleri[1,3]

[1] Max Planck Institute for the Structure and Dynamics of Matter, 22761 Hamburg, Germany

[2] The Harrison M. Randall Laboratory of Physics, University of Michigan, Ann Arbor, Michigan, USA

[3] Department of Physics, Clarendon Laboratory, University of Oxford, Oxford OX1 3PU, United Kingdom

e-mail: michele.buzzi@mpsd.mpg.de, andrea.cavalleri@mpsd.mpg.de

**The optical excitation of circular phonons has opened up non-equilibrium phenomena with no equilibrium counterparts, ranging from the generation of synthetic magnetic fields to the control of topological order or the manipulation of ferroaxial states. In this area, significant controversy has arisen on whether sizeable "true" magnetizations accompany circular phonons. Several confounding factors cloud the interpretation of the experiments, which are based on probing polarization rotation in the driven material. These range from the interpretation of the Faraday effect in non-equilibrium conditions to spurious nonlinear optical contributions in pump probe experiments. Here we resolve these ambiguities by measuring both polarization rotation inside the driven material and the stray magnetic field generated outside it with a proximate Faraday detector. For a 5 MV/cm mid-infrared drive, we observe a colossal polarization rotation inside a driven SiC crystal, similar to what was reported in previous work. Depending on the assumptions made, this could correspond to magnetizations as high as $\mu_0 M =$ 0.3 T or as low as 4 μT. However, when measuring the magnetic stray field outside the driven sample we unambiguously determine that the magnetization in SiC is negligibly small ($\mu_0 M < 0.001$ T).**

Ultrashort laser pulses at THz and mid-infrared frequencies can drive collective modes to large amplitudes and manipulate material properties at ultrafast timescales. In particular, coherently driven circular phonons break time-reversal symmetry and can trigger magnetic responses in magnetic[1–3] and non-magnetic materials[4–7]. Circular phonon excitation has also been shown to induce topological effects[8,9] and helicity-dependent switching of ferro-axial order[10].

In an ionic solid, the circular motion of the ions generates microscopic current loops and local orbital magnetic moments (Figure 1a). However, owing to the large ionic masses and the slow angular velocities, these moments are expected to be small even under strong excitation ($\sim 10^{-4}\ \mu_B$)[11].

In $SrTiO_3$, large circular-phonon-induced Faraday rotation was interpreted as evidence for magnetizations that are four orders of magnitude larger than expected from ionic currents alone[4]. These observations have attracted considerable theoretical interest[12,13], and triggered similar studies in other ferroelectrics[5,6].

However, an important source of uncertainty in these measurements descends from the fact that the antisymmetric part of the third order non-linear optical susceptibility $\chi^{(3)}$ can produce large anomalous Faraday rotations that are unrelated to magnetism[14–17] (Figure 1a, Supplementary Section S1). These nonlinear optical responses generate a pump-induced circular birefringence and reverse sign with the helicity of the pump, making them indistinguishable from magnetic Faraday rotation. In addition, residual linearly polarized pump components can complicate the interpretation[5].

Polarization rotation measured within the material's photoexcited volume therefore cannot, by itself, unambiguously determine the magnitude of the phonon-induced magnetization. Yet, while an induced magnetization generates a magnetic stray field that extends outside the photo-excited volume of the material (Figure 1b), spurious effects related to nonlinear mixing do not. Hence, by measuring this stray magnetic field one can test whether the large transient magneto-optical responses are accompanied by a large magnetization.

Here we report an ultrafast stray field magnetometry measurement in 6H-SiC, an insulating polar semiconductor for which circular excitation of a doubly degenerate $E_1$ phonon mode is expected to generate a magnetization[11]. In our experiments, this mode was driven with circularly polarized mid-infrared pulses, combined both with magneto-optical probing inside SiC and with an independent measurement of the stray magnetic field in a separate detector placed outside the driven solid[18,19].

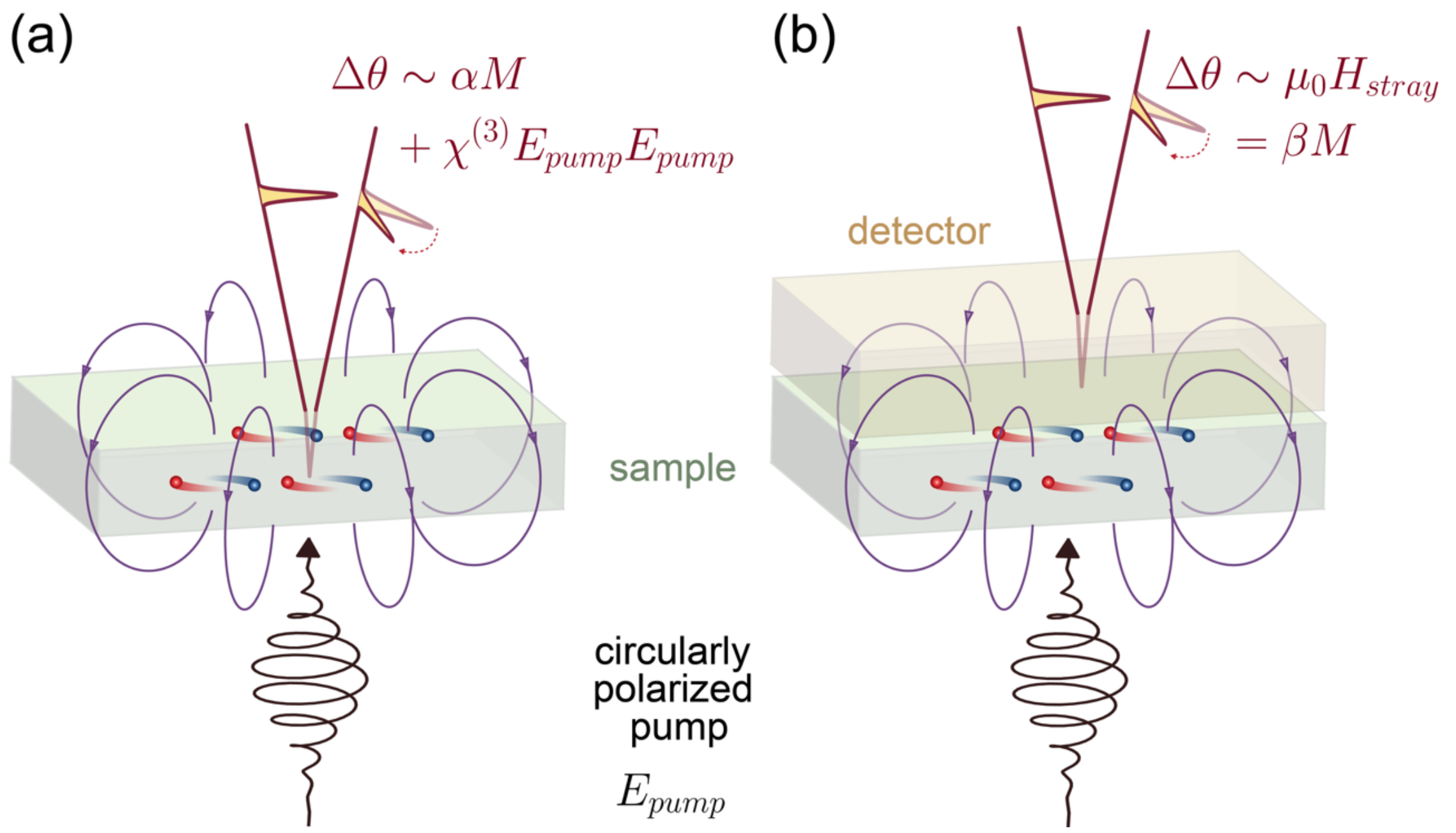

**Figure 1 | Probing the magnetization induced by circular phonons. (a)** Circularly polarized light induces a circular motion of the ions in the material (green slab), which act as current loops, and generate a magnetic moment. Polarization rotation is measured directly in the photo-excited volume. The signal consists of two contributions: one proportional through a coefficient $\alpha$ to the induced magnetization through the Faraday effect, and a second one due to a four-wave mixing process between the pump and the probe. **(b)** The magnetic stray field generated by the phonon induced magnetization is measured using a magneto-optic detector (orange slab). The Faraday rotation accumulated in the detector slab is proportional to the sensed stray field which is directly proportional through a geometry-determined calibration factor $\beta$ (Supplementary Section S8) to the phonon-induced magnetization.

## Helicity Dependent Polarization Rotation inside 6H-SiC

The crystal structure of 6H-SiC is shown in Figure 2a. The doubly degenerate $E_1$ phonon mode targeted in this work supports the circular motion of the Si and C atoms with the same handedness in the ab-plane (Figure 2b). The equilibrium linear reflectivity spectrum measured by Fourier Transform Infrared spectroscopy is shown in Figure 2c alongside a Lorentz model fit. The measured spectrum agrees well with existing literature[20] and exhibits an approximately 6 THz-wide Reststrahlen band extending from $\Omega_{TO} = 23.9$ THz to $\Omega_{LO} = 29.8$ THz, reflecting the strong infrared activity of this mode (Supplementary Section S2).

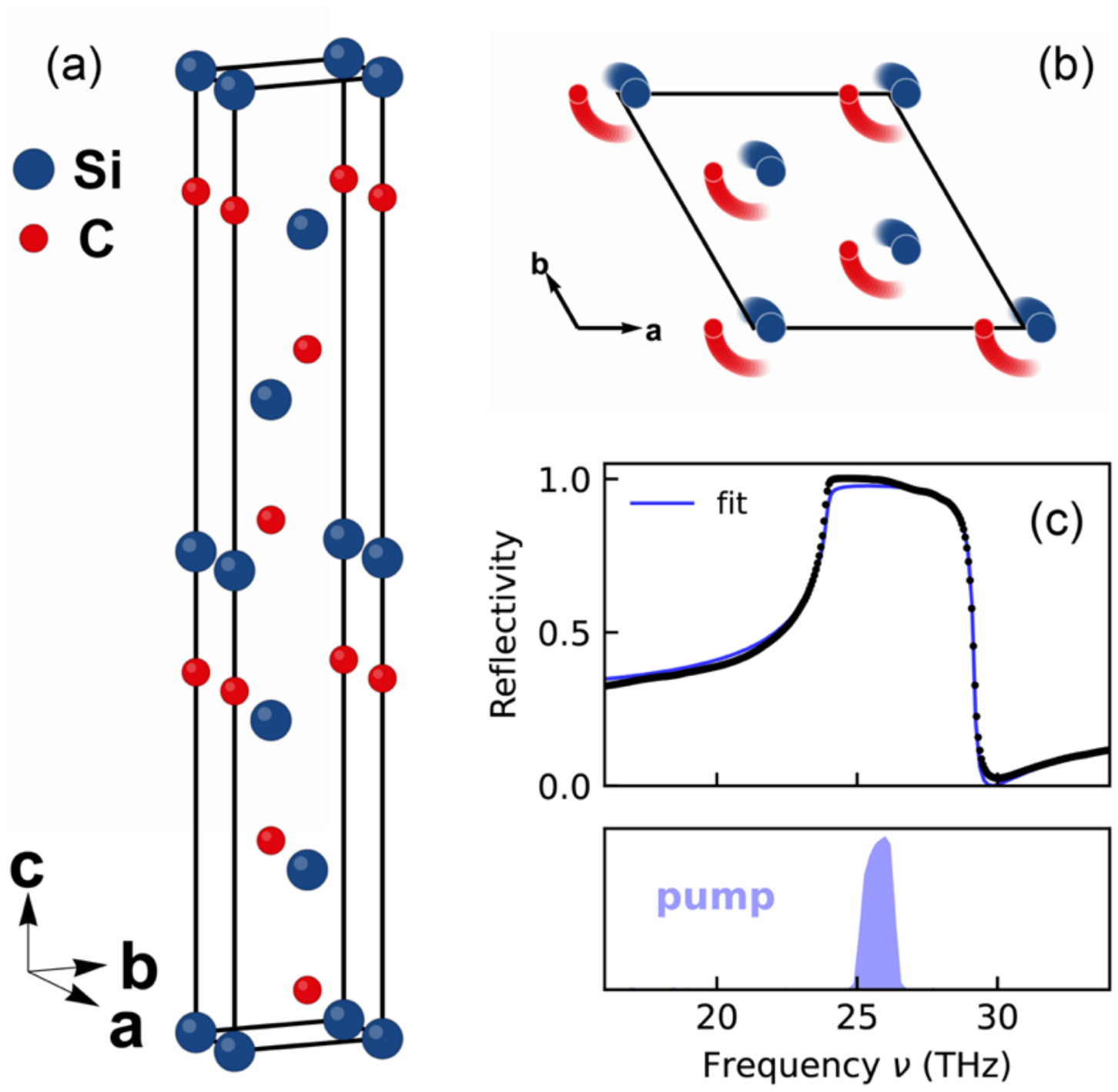

**Figure 2 | Circular phonons in 6H-SiC. (a)** Unit cell of 6H-SiC. **(b)** The two degenerate $E_1$ modes in the ab-plane are driven $\pi/2$ out of phase, generating a circular motion of the Si and C ions. **(c)** Frequency dependent reflectivity of 6H-SiC measured with Fourier Transform Infrared Spectroscopy (FTIR) with light polarized in the ab-plane showing the Reststrahlen band of the $E_1$ modes. The panel below indicates the measured spectrum of the pump pulses used to excite the circular phonons.

We excited this phonon mode using circularly polarized, ~500 fs long, mid-infrared pulses centered at 25.8 THz (Figure 2c). These were generated by difference frequency mixing of signal and idler pulses from an optical parametric amplifier (Supplementary Section S3). The linearly polarized pump pulses were converted to circularly polarized using an interferometer-based variable waveplate. Independent control of its retardance and effective optical axis allowed us to compensate for polarization effects from the metallic steering mirrors, and ensure better than 95% polarization purity at the sample position (Supplementary Section S4).

The polarization rotation in the photoexcited volume was measured using a time-delayed, linearly polarized 800 nm probe pulse. As shown in Figure 3a, the pump and the probe pulses reached the 6H-SiC sample from opposite sides at near normal incidence. To distill Faraday-like polarization rotation signals that are non-reciprocal and independent of the incoming probe polarization we acquired the data as follows. First, the polarization of the probe pulse was analyzed after being reflected from the second surface of the sample, i.e. traversing the photo-excited material twice. Second, we varied the probe polarization angle $\theta_{inc}$ from 0 to $2\pi$ and averaged the measured rotation over these angles to extract the polarization independent component, suppressing contributions from pump-induced linear birefringence and dichroism (Supplementary Section S1, S5). Finally, we isolate the helicity-dependent response by taking the half-difference between signals acquired with opposite pump helicities. The measured data are shown in Figure 3b alongside an exponential rise and decay fit. Upon photo-excitation a time-dependent polarization rotation signal was observed, peaking at ~3 μrad. The signal shows a fast onset and a slower decay, with a full width at half maximum of ~700 fs, slightly longer than the ~500 fs pump pulse (Figure 3b), possibly due to the finite lifetime of the driven $E_1$ phonon. Beyond the anomalous Faraday rotation generated via four-wave mixing processes discussed above, which could in principle be calibrated and subtracted away[4], the interpretation of these data is intrinsically ambiguous. Indeed, if one attributes the observed polarization rotation only to the magnetic Faraday effect and starts from the equilibrium Verdet constant of 6H-SiC (Supplementary Section S1), one extracts a response equal in magnitude to that achieved by a large equivalent magnetic field $B_{eq} \approx 0.3\,\mathrm{T}$. One could then conclude that in absence of an applied magnetic field $H$, the polarization rotation derives from an optically induced magnetic induction $B \approx 0.3$ T, and hence from an optically induced magnetization $\mu_0 M \approx 0.3$ T, as done in Ref. [4].

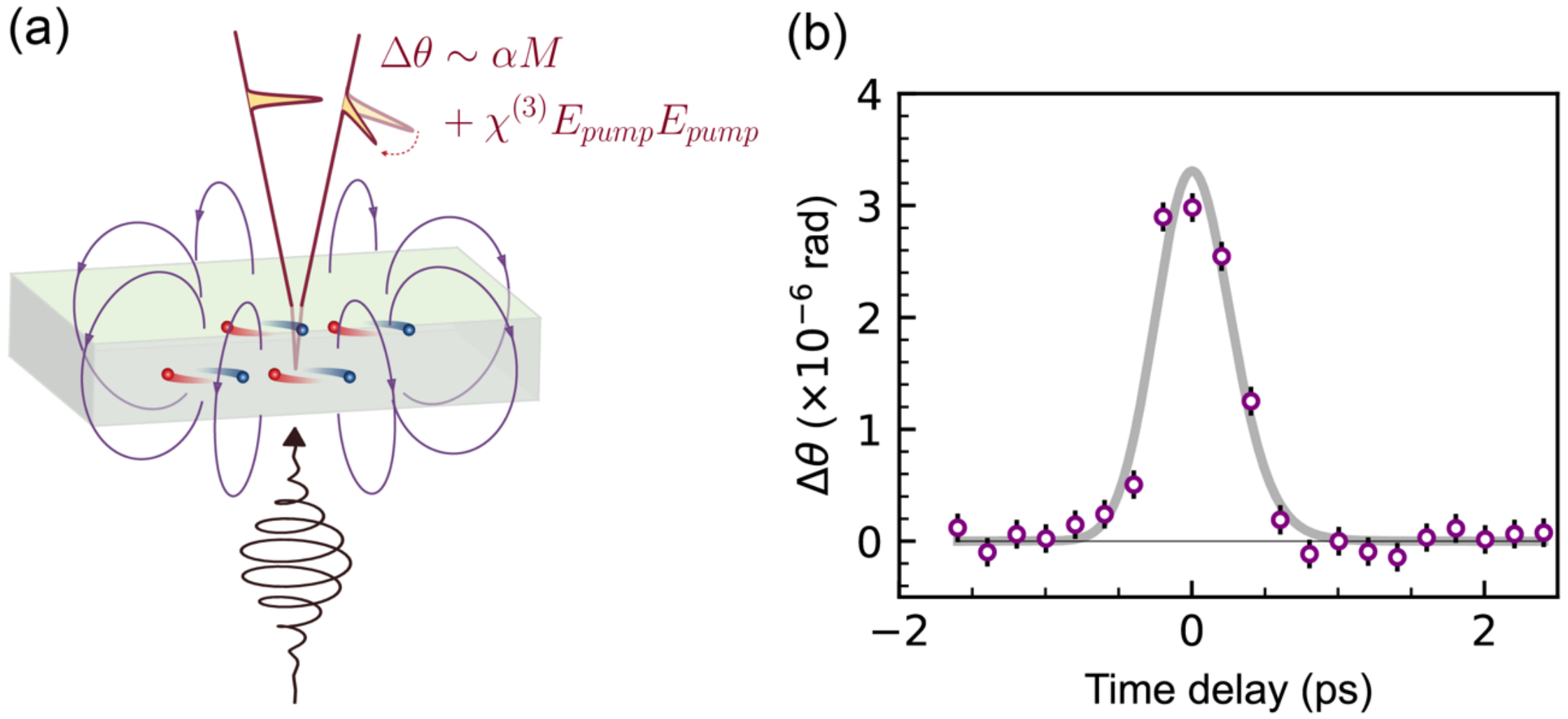

**Figure 3 | Polarization rotation measurement in the photo-excited volume. (a)** Schematic of the experiment. A 35 µm-thick 6H-SiC crystal (green slab) is excited with circularly polarized mid-infrared pulses resonant with the doubly-degenerate $E_1$ phonon mode. The polarization rotation of a time-delayed probe pulse is measured within the photo-excited volume and can contain contributions both from the magnetisation-induced Faraday effect and nonlinear optical four-wave mixing between the pump and the probe. **(b)** Time-dependent polarization rotation, obtained as the half-difference between signals measured for opposite pump-pulse helicities (↻ and ↺). The crystal c-axis, oriented vertically, defines the positive direction of photon angular momentum. The symbols ↻ and ↺ indicate the direction of the temporal rotation of the electric field within the crystal, viewed along the positive c-axis. These data were acquired at room temperature, using ~500 fs long, 25.8 THz center-frequency circularly polarized pump pulses with a fluence of 10 mJ/cm$^2$, above the saturation fluence threshold. A fluence dependence is reported in Supplementary Section S9.

On the other hand, if one assumes instead that the driven phonons should be considered as an *externally* applied field $\mu_0 H = B_{eq}$, as done in Refs. [2,5,21], one would extract a magnetization $\mu_0 M \approx |\chi_{SiC}| B_{eq} \approx 4$ µT ($\chi_{SiC} = -1.3 \times 10^{-5}$ is the equilibrium volume magnetic susceptibility of 6H-SiC). Note also that it is not certain that the magnetization generated by a driven circular phonon, which couples to electronic degrees of freedom in different ways from an applied magnetic field, should then be extracted from Faraday rotation using the same constants calibrated in equilibrium with DC magnetic fields. Hence, it is very doubtful that polarization rotation in the driven solid can be used *at all* to establish the magnitude of the phonon-induced magnetization.

## Probing the magnetic stray field using ultrafast magnetometry

To clarify these issues we combined the measurements discussed above with measurements of the magnetic stray field in the vicinity of the photo-excited volume of the 6H-SiC sample, similar to what done previously in photo-excited superconductors[18,19].

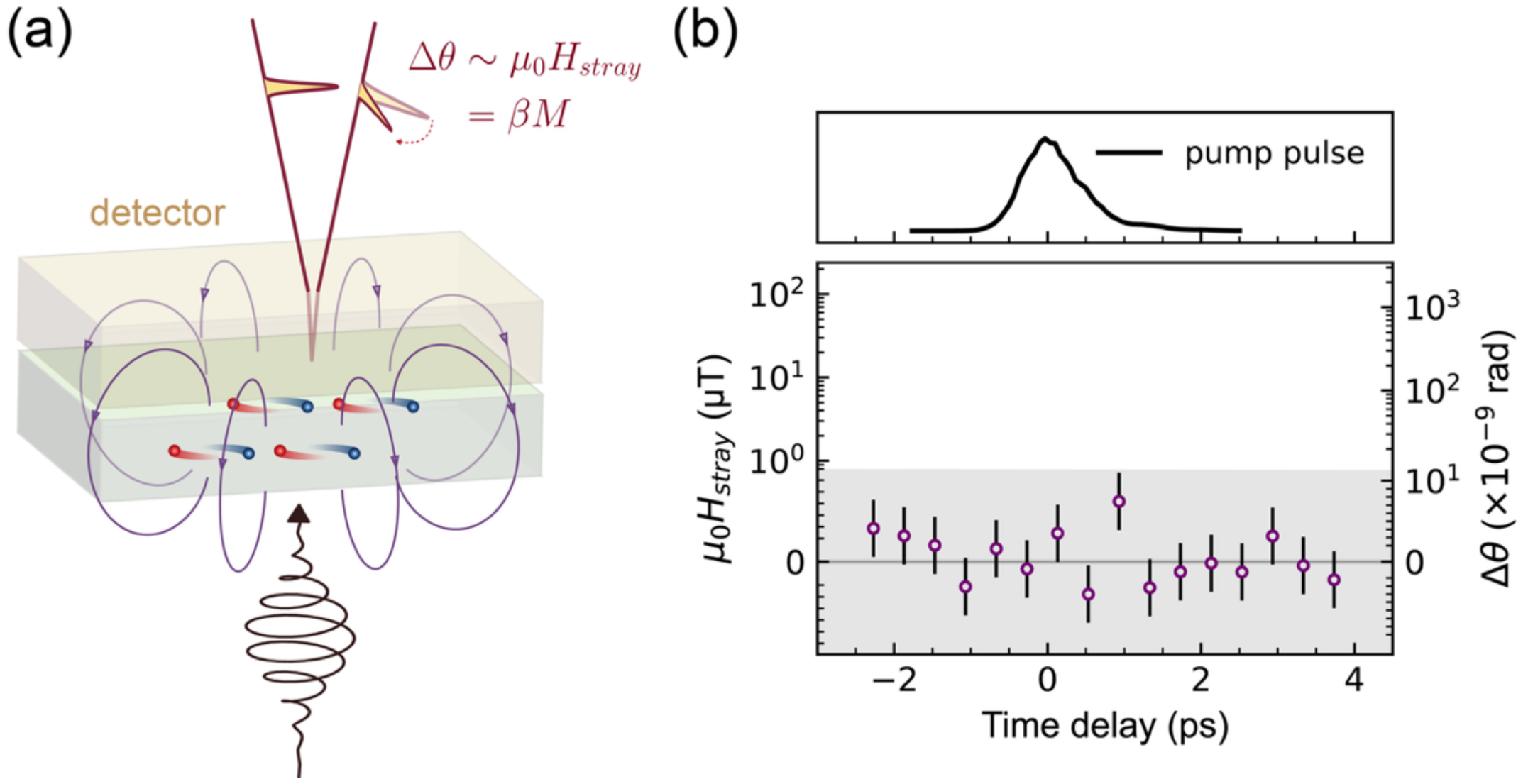

**Figure 4 | Stray magnetic-field measurement. (a)** Schematic of the experiment. A 35 μm-thick 6H-SiC crystal (green slab) is excited with circularly polarized mid-infrared pulses resonant with the doubly-degenerate $E_1$ phonon mode. A 60 μm thick GaP crystal (orange slab) is placed in the vicinity of the unpumped surface of the 6H-SiC crystal to measure the stray field generated by circular-phonon-induced magnetization. **(b)** Time-dependent polarization rotation measured in the GaP detection crystal, obtained as the half-difference between signals measured for opposite pump-pulse helicities (↻ and ↺). The grey band represents the +/- 800 nT uncertainty band of the experiment. These data were acquired at room temperature, using ~500 fs long, 25.8 THz center-frequency, circularly polarized pump pulses with a fluence of ~40 mJ/cm$^2$.

A schematic of the experiment is shown in Figure 4a. A ~60 μm thick GaP (100) detector was placed above a 35 μm thick 6H-SiC crystal. The SiC sample was photoexcited from the side opposite the GaP detector with circular polarized pump pulses. A time-delayed 800 nm probe pulse measured the Faraday rotation in the GaP crystal, yielding a local measurement of the magnetic stray field surrounding the SiC photo-excited volume. The pump spectrum lay entirely within the Reststrahlen band and the SiC sample thickness was sufficient to suppress the pump transmission to the GaP detector. The SiC crystal was polished flat, whereas the GaP detector was wedged and positioned at a slight angle such that only the reflection from the GaP surface facing the SiC sample was collected and guided into the polarization analyzer, isolating the polarization rotation accumulated during the double pass through the GaP detector (Supplementary Section S6).

Figure 4b shows the time-dependent magnetic stray field measurement in the GaP detector together with a cross-correlation measurement of the pump intensity profile in a bare area of the GaP detector. No pump-induced magnetic stray field was resolved within an experimental uncertainty of better than 1 μT, corresponding to a Faraday rotation of 10 nrad for the specific detector thickness considered here (Supplementary Section S7).

This result sets a very low upper bound on the true magnetization generated by circular-phonon excitation in 6H-SiC. To quantify this bound, we used finite-element simulations to calculate the time-dependent magnetic stray field surrounding the photoexcited volume (Supplementary Section S8). The expected value of the magnetic stray field detected by the GaP was calculated averaging the component of the magnetic stray field along the probe propagation direction over the forward and return paths inside the detector crystal. The geometry for the finite-element simulations is shown in Figure 5a. A ~60 μm thick GaP (100) detector was placed 5 μm above a 35 μm thick 6H-SiC crystal. The photo-induced magnetization was assumed to be spatially homogeneous within a thin disk whose diameter matched that of the pump beam and whose thickness corresponded to the pump penetration depth. The magnetization was assumed to point along the pump propagation direction and to follow an exponential rise and decay profile reproducing the polarization-rotation dynamics measured in SiC (inset in Figure 5a). A snapshot of the calculated magnetic stray field profile at the peak of the response is shown in Figure 5a.

Repeating these simulations for different magnetization peak values allowed us to calculate a calibration curve, shown in Figure 5b, relating the peak polarization rotation in GaP to the peak magnetization value in the photoexcited volume in SiC.

The measured data are shown after calibration in Figure 5c on a symmetric logarithmic plot. The experimental uncertainty sets an upper bound for the phonon-induced magnetization of $\mu_0 M < 1\ \mathrm{mT}$. For comparison the expected response for $\mu_0 M_{peak} = 300\ \mathrm{mT}$ is shown as a dashed line. The experiments reported here exclude a transient magnetization of this magnitude.

**Discussion and Conclusions**

Taken together, the measurements discussed above demonstrate that the observation of a large helicity-dependent polarization rotation in an insulator does not imply a comparably large magnetization, validating the original theoretical estimates. We determine an upper bound on the phonon-induced magnetization of $\mu_0 M < 1\ \mathrm{mT}$. This bound is more than two orders of magnitude smaller than what one would conclude by estimating an induced magnetization generating a magnetic induction equal to the field $B_{eq}$. More broadly, ultrafast stray-field magnetometry provides an experimental basis to extract macroscopic magnetization in driven and non-equilibrium quantum solids, bypassing many of the controversial interpretations that have plagued the field of optomagnetism. In the future, wide application of this method will likely be extended to other experimental platforms, placing ultrafast photo-magnetic phenomena on a firm experimental footing.

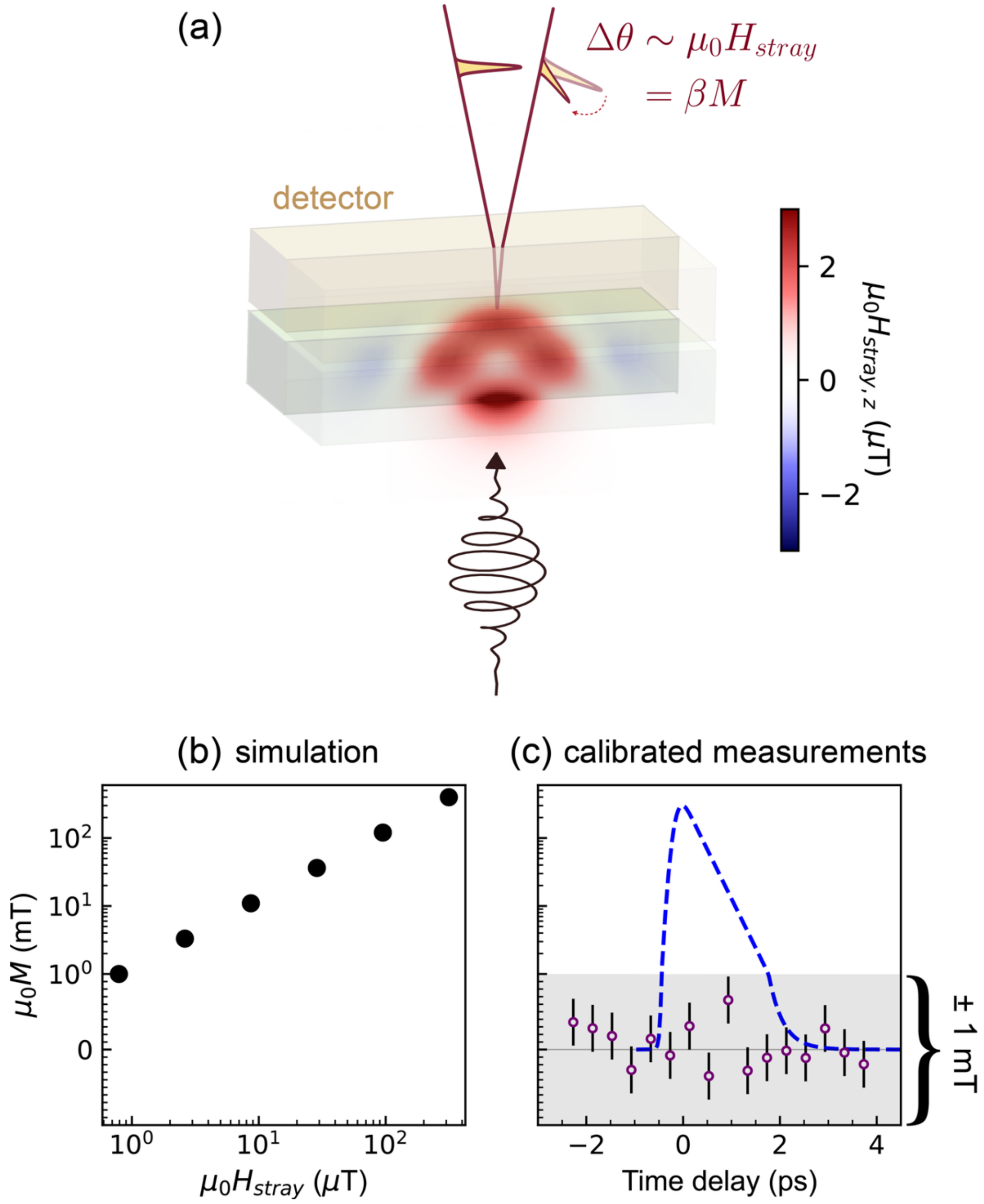

**Figure 5 | Stray magnetic-field measurement and calibration. (a)** Schematic of the experiment and geometry used to calculate the time-dependent stray magnetic field. The photoexcited 6H-SiC region is modelled as a magnetized disk with a diameter matching the pump beam and a thickness corresponding to the pump penetration depth. The magnetization is assumed to follow the temporal profile of the pump-induced polarization rotation dynamics shown in Fig. 3b. The colored shading represents the calculated spatially dependent z-component of the magnetic field at the peak of the response, obtained by integrating Maxwell's equations (Supplementary Section S8). **(b)** Numerical calibration relating the assumed peak magnetization in 6H-SiC to the expected peak Faraday rotation generated in the GaP detector. **(c)** Time-dependent polarization rotation measured in the GaP detector upon excitation of 6H-SiC, converted into an equivalent magnetization using the calibration in (b). The grey shaded region indicates an experimental upper bound of approximately ±1 mT on the phonon-induced magnetization. The blue dashed curve represents the expected signal for a peak magnetization of 300 mT.

## Acknowledgments

We acknowledge support from the Deutsche Forschungsgemeinschaft (DFG) via the Cluster of Excellence CUI: Advancing Imaging of Matter (EXC 2056, project ID 390715994). We are also grateful to Kilian Keller for helping with setting up the mid-infrared parametric amplifier, Hangoo Lee for help with device fabrication, Boris Fiedler, Birger Höhling and Toru Matsuyama for their support in the fabrication of the electronic devices used on the measurement setup, to Elena König, Guido Meier, and Daniele Nicoletti for help with sample characterization. Discussions with Lara Benfatto and Niccolò Sellati are gratefully acknowledged.

## Competing Interests

The authors declare no competing interests.

## Data Availability

Source data are provided with this paper. All other data that support the plots within this paper and other findings of this study are available from the corresponding authors.

# Circular phonons generate negligible magnetization in insulators

C-C. Wu[1], M. Buzzi[1], A. Iudica[1], R. Romero III[1], G. De Vecchi[1], R. Merlin[1,2], A. Cavalleri[1,3]

[1] Max Planck Institute for the Structure and Dynamics of Matter, 22761 Hamburg, Germany

[2] The Harrison M. Randall Laboratory of Physics, University of Michigan, Ann Arbor, Michigan, USA

[3] Department of Physics, Clarendon Laboratory, University of Oxford, Oxford OX1 3PU, United Kingdom

e-mail: michele.buzzi@mpsd.mpg.de, andrea.cavalleri@mpsd.mpg.de

## Supplementary Material

# S1. Four-wave Mixing Contributions to Polarization Rotation

## S1.1 – Pump-induced Circular Birefringence

As discussed in the main text, when SiC is driven with circularly polarized laser pulses, four wave mixing between the pump and the probe can generate an anomalous polarization rotation in the absence of a magnetization. In this section we analyze how $\chi^{(3)}$ effects enter the polarization rotation measurements.

Let us denote the dielectric tensor of the pumped material in the ab-plane by $\epsilon_{ij}(\omega)$, and by $\epsilon_{ij}^{eq}(\omega)$ the dielectric tensor at equilibrium. $\chi_{ijkl}^{(3)}$ is the third-order nonlinear susceptibility tensor, $\omega$ is the probe frequency in the optical range, and $\Omega$ is the pump frequency in the mid-infrared range. The four-wave mixing process $\omega = \omega - \Omega + \Omega$ gives rise to emission of frequency identical to the input probe frequency; in other words, this process acts as an effective change to the linear dielectric tensor. This can be formalized as follows:

$$\epsilon_{ij}(\omega) = \epsilon_{ij}^{eq}(\omega) + \chi_{ijkl}^{(3)}(\omega; \omega, -\Omega, \Omega) E_k(-\Omega) E_l(\Omega) \tag{S1.1}$$

For a circularly polarized pump:

$$E(\Omega) = E_0 \begin{pmatrix} 1 \\ -i \end{pmatrix} \tag{S1.2}$$

$$E(-\Omega) = E(\Omega)^* \tag{S1.3}$$

in the *6mm* point group (appropriate for 6H-SiC), before applying Kleinman symmetry, one obtains:

$$\epsilon_{ij}(\omega) = \begin{pmatrix} \epsilon_{11}^{eq} & 0 \\ 0 & \epsilon_{11}^{eq} \end{pmatrix} + E_0^2 \begin{pmatrix} \chi_{1111} + \chi_{1122} & -i(\chi_{1212} - \chi_{1221}) \\ i(\chi_{1212} - \chi_{1221}) & \chi_{1111} + \chi_{1122} \end{pmatrix} \tag{S1.4}$$

$$\Delta\epsilon_{ij} = E_0^2 \begin{pmatrix} \chi_{1111} + \chi_{1122} & -i(\chi_{1212} - \chi_{1221}) \\ i(\chi_{1212} - \chi_{1221}) & \chi_{1111} + \chi_{1122} \end{pmatrix} \tag{S1.5}$$

The imaginary off-diagonal terms in red give rise to a circular birefringence similar to the Faraday effect. If Kleinman symmetry holds in addition, i.e. when the susceptibility is dispersionless, one obtains that $\chi_{1212} = \chi_{1221}$, the off-diagonal components of $\epsilon_{ij}(\omega)$ vanish and the circular pump does not introduce circular birefringence. However, as pointed out in Refs. [1–3], near phonon resonances the dispersion of the susceptibility cannot be neglected, leading to a breakdown of Kleinman symmetry and allow a circularly polarized pump pulse to introduce circular birefringence. This change in off-diagonal dielectric tensor $\Delta\epsilon_{ij}$ is proportional to $E_0^2$ and reverses sign when pump helicity changes.

For clarity we extract in the following the form of the dielectric tensor of a medium magnetized by the pump. The Faraday effect arises from a magneto-optic coupling of type $\Delta F \sim \chi_{ijk}^{MO} E_i E_j M_k$, where $F$ is the free energy of the system, $E_{i,j}$ are the components of the probe electric field and $M_k$ is the $k$ component of the pump-induced magnetization. Since $M_k$ is an axial and time-reversal-odd vector, $\chi_{ijk}^{MO} E_i^* E_j$ must also be an axial and time-reversal-odd vector to make the energy a polar and time-reversal-even scalar. The only way for the coefficient tensor $\chi_{ijk}^{MO}$ to achieve that is $\chi_{ijk}^{MO} E_i^* E_j = i\chi^{MO}\big(E^*(\omega) \times E(\omega)\big)_k$, up to a constant factor. Namely, $\chi_{ijk}^{MO}$ is antisymmetric in the $i, j$ indicies. From this, we can extract the change in the dielectric tensor as follows:

$$P_i \equiv -\frac{\partial F}{\partial E_i^*} = \chi_{ijk}^{MO} E_j M_k \tag{S1.6}$$

$$\Delta\epsilon_{ij} = \frac{\partial P_i}{\partial E_j} = \chi_{ijk}^{MO} M_k = \begin{pmatrix} & i\chi^{MO} M_3 \\ -i\chi^{MO} M_3 & \end{pmatrix} \tag{S1.7}$$

A comparison of the $\Delta\epsilon_{ij}$ associated with the four-wave mixing signal with circular pump and that associated with a "true" Faraday effect shows that from a symmetry standpoint they will both result in a polarization rotation of a linear probe, that is independent of the incoming polarization angle. Furthermore, the same argument for the magneto-optic coupling holds also for the pump electric field, i.e., at the pump frequency $\big(E^*(\Omega) \times E(\Omega)\big)_k$ constitutes the conjugate field of magnetization $M_k$, which implies that the sign the induced $M_k$ reverses with the pump helicity, and that the amplitude of the induced $M_k$ is proportional to $E_0^2$. Hence, neither the fluence nor the helicity dependence can discriminate between the two effects.

### S1.2 – Pump-induced Linear Birefringence

The same analysis for linear pump polarization in the *6mm* point group shows that pumping induces a linear birefringence term whose axis is determined by the pump linear polarization. For a vertically polarized pump:

$$E(\Omega) = E_0 \begin{pmatrix} 1 \\ 0 \end{pmatrix} \tag{S1.8}$$

$$E(-\Omega) = E(\Omega)^* \tag{S1.9}$$

The $\chi_{ijkl}^{(3)}$ effect manifests as an induced linear birefringence for a linearly polarized probe:

$$\epsilon_{ij}(\omega) = \epsilon_{ij}^{eq}(\omega) + \chi_{ijkl}^{(3)}(\omega;\omega,-\Omega,\Omega)E_k(-\Omega)E_l(\Omega) =$$
$$= \begin{pmatrix} \epsilon_{11}^{eq} & 0 \\ 0 & \epsilon_{11}^{eq} \end{pmatrix} + E_0^2 \begin{pmatrix} \chi_{1111} & 0 \\ 0 & \chi_{1122} \end{pmatrix} \tag{S1.10}$$

For arbitrary pump polarizations in the ab-plane of a hexagonal lattice, due to its isotropy, one finds that a linearly polarized pump always introduces in $\epsilon_{ij}(\omega)$ a term $\chi_{1111}E_0^2$ in the direction of the polarization and $\chi_{1122}E_0^2$ perpendicular to it. As such, for a linearly polarized probe at an angle $\theta$ away from the pump polarization, a phase retardation $\phi \propto (\chi_{1111} - \chi_{1122})E_0^2$ is introduced between the probe polarization components along the pump-induced principal axes:

$$E_{probe}^{in} = \begin{pmatrix} \cos\theta \\ \sin\theta \end{pmatrix} \rightarrow E_{probe}^{out} = \begin{pmatrix} \cos\theta \\ \sin\theta\, e^{i\phi} \end{pmatrix} \tag{S1.11}$$

This results in an induced-ellipticity, which for $\phi \ll 1$ becomes:

$$\eta = \frac{1}{2}\arcsin\left(\frac{2\Im[E_x^* E_y]}{|E|^2}\right) = \frac{1}{2}\arcsin(\sin 2\theta \sin\phi) \sim \phi \propto (\chi_{1111} - \chi_{1122})E_0^2. \tag{S1.12}$$

## S1.3 – Polarimetry of the pump-induced components to the signal

Using Jones calculus and considering the dielectric tensors extracted above, one can predict how the signal from the different contribution would depend on the incoming probe polarization through balanced detection with a half waveplate (Supplementary Section

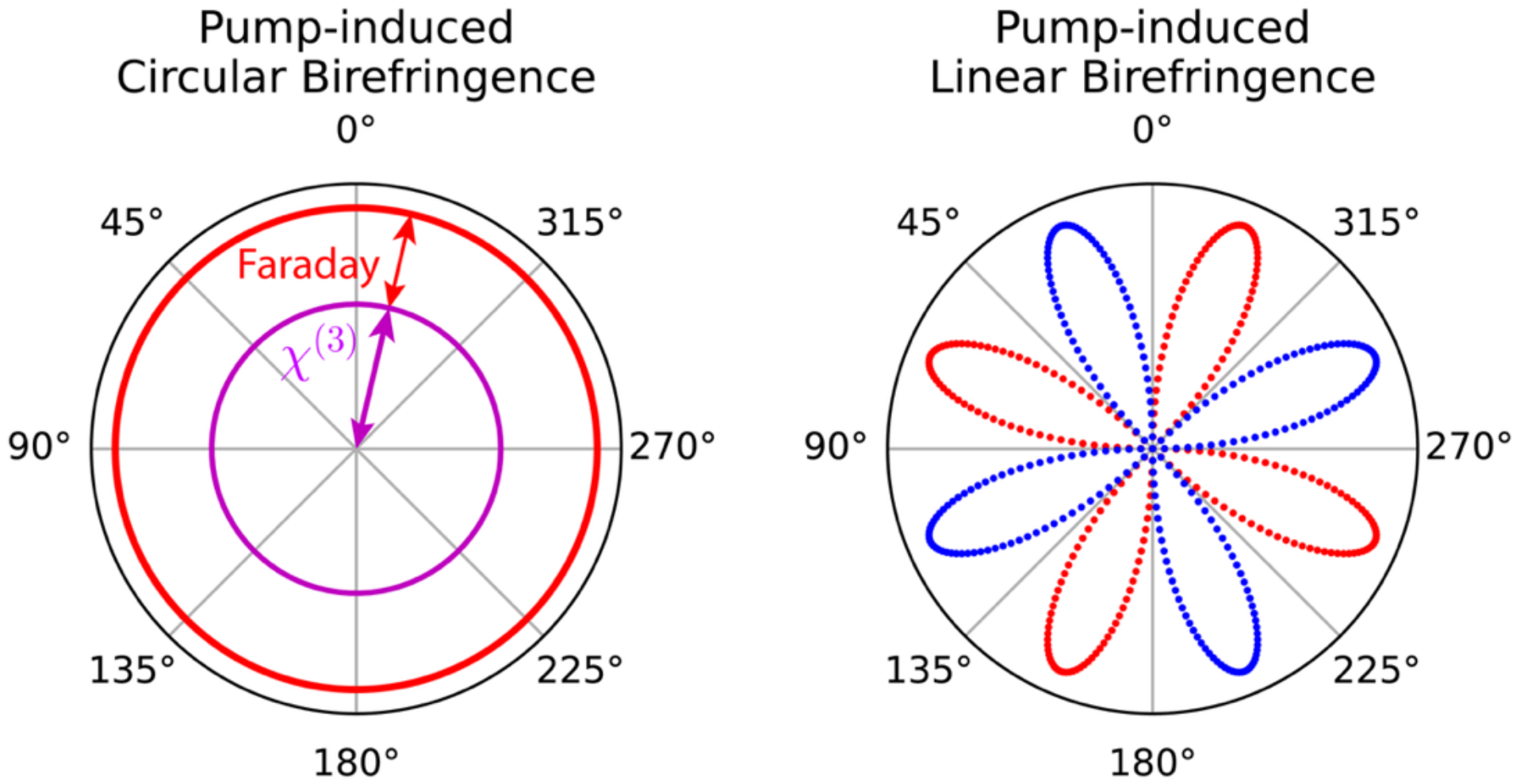

**Figure S1.1. Polarimetry of the induced change in dielectric tensor.** The polar plots show the calculated polarization rotation at a given incoming probe polarization for **(a)** pump-induced circular birefringence and **(b)** linear birefringence. The blue and red data points in (b) are used to indicate polarization rotation in opposite directions.

S3). As shown in Figure S1.1a, both the circular birefringence induced by the Faraday effect and that induced by $\chi^{(3)}$ are independent of the incoming probe polarization. The pump-induced linear birefringence (Figure S1.1b), instead, shows a 4-fold $\sin 4\theta$ dependence. Hence, performing a measurement averaging over all probe polarization angles eliminates pump-induced linear birefringence and isolates the circular birefringence contribution.

## S2. Sample Preparation and Characterization

The undoped 6H-SiC single crystals were obtained through a commercial supplier (SurfaceNet GmbH). The crystals, of size 5x5x0.25 mm$^3$ were cut to expose an ab-oriented surface. Their orientation was determined by the manufacturer using Laue x-ray diffraction.

Their reflectivity in the mid-infrared range, reported in Figure 2 of the main text, was measured using a Fourier Transform Infrared Spectrometer (Bruker Vertex 80) using a Globar source, Ge coated KBr beamsplitter and a Mercury Cadmium Telluride detector.

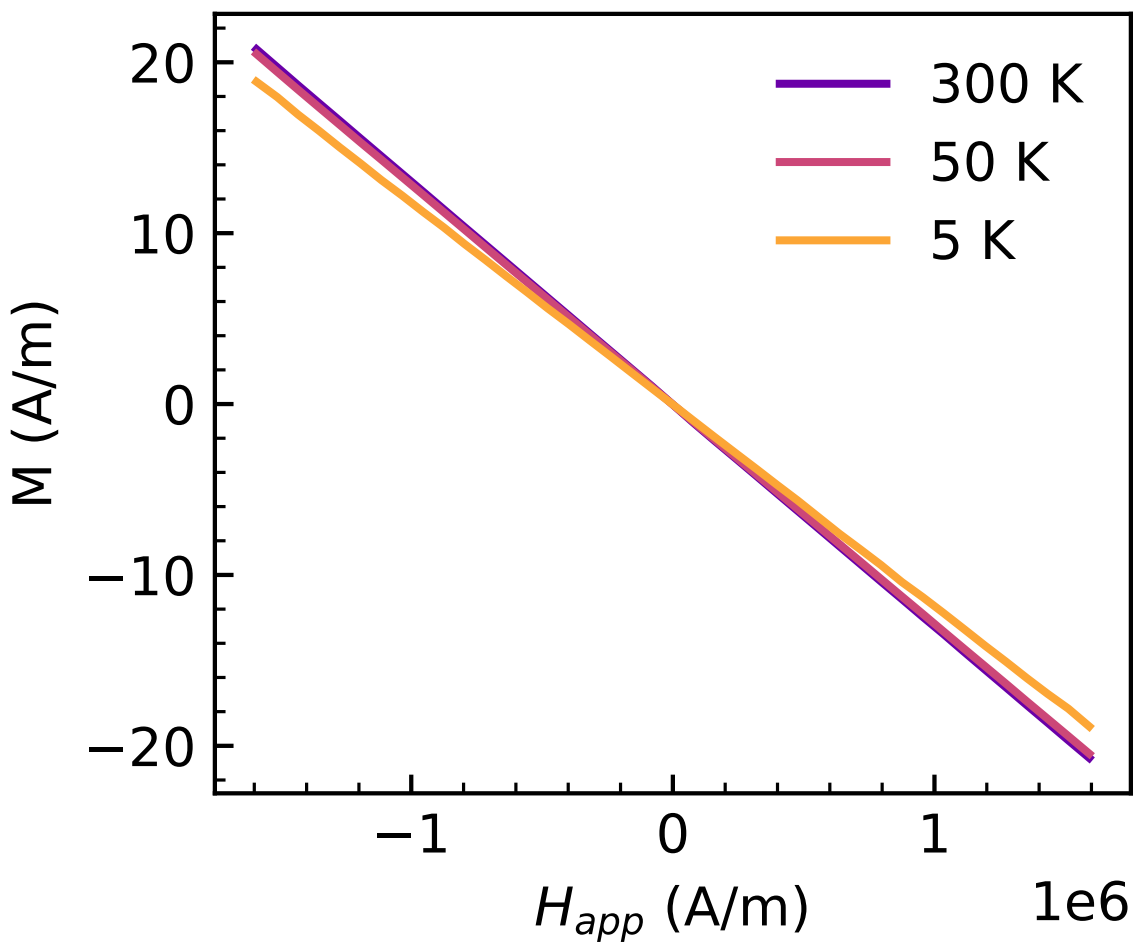

**Figure S2.1.** Field dependent DC magnetization measurement. The measurements were performed in a SQuID magnetometer with the magnetic field applied along the crystal c-axis. The extracted value of the room-temperature volume magnetic susceptibility is $\chi_{SiC} = -1.3 \times 10^{-5}$ at room temperature.

To characterize their diamagnetic susceptibility, we carried out DC magnetization measurements in a Quantum Design MPMS3 SQuID magnetometer as a function of the applied field along the c-axis (Figure S2.1) at room temperature. The sample shows a clear diamagnetic behavior, indicating the absence of unwanted magnetic impurities. A linear fit to the data yielded the value of the room-temperature volume diamagnetic susceptibility of 6H-SiC, $\chi_{SiC} = -1.3 \times 10^{-5}$.

The magneto-optic response of 6H-SiC was characterized measuring the polarization rotation of an 800 nm laser pulse propagating along the sample c-axis for different values of the applied magnetic field along the beam propagation direction. The Faraday effect directly relates the magnetic field applied to a material to the polarization rotation as follows:

$$\theta = VBL \tag{S2.1}$$

where θ represents the rotation of the polarization of the input beam, B is the magnitude of the applied magnetic field along the light propagation direction inside the medium and L is the thickness of the medium. The proportionality constant V is known as the Verdet constant, which is a material dependent constant depending also on other parameters such as the wavelength of the incoming polarized light. The measured data is reported in Figure S2.2. A linear fit using the Faraday effect relation yields a value for the Verdet constant of 6H-SiC of 19 rad $T^{-1}m^{-1}$ at 800 nm wavelength.

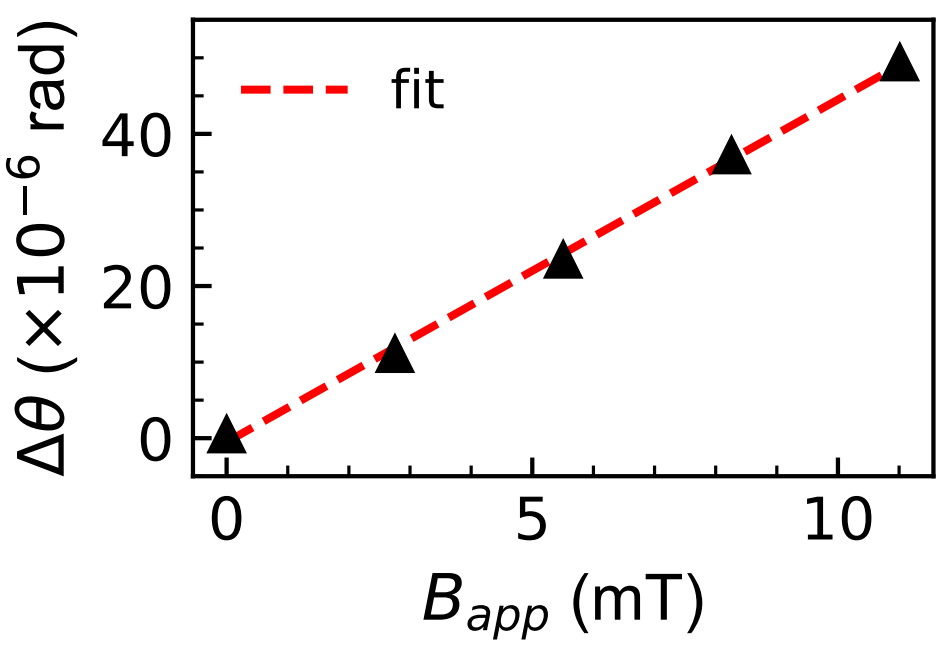

**Figure S2.2.** Field dependent polarization rotation measured in transmission using a 240 μm thick 6H-SiC sample at room temperature. The extracted value of the Verdet constant is 19 rad/T·m.

## S3. Experimental Setups and Data Acquisition

### S3.1 – Experimental Setup

The measurements were performed in the setup shown in Figure S3.1. Two commercial Ti:Sapphire regenerative amplifiers (Spectra-Physics Spitfire ACE) were seeded by the same Ti:Sapphire oscillator, and generated pulses at 1 kHz repetition rate (pulse duration 100 fs, pulse energy 6 mJ) and 2 kHz repetition rate (pulse duration 35 fs, pulse energy 2.5 mJ) respectively. The probe pulses were directly derived from the fundamental output of the 2 kHz amplifier. The mid-infrared pump pulses (MIR) were obtained from difference frequency generation (DFG) in GaSe between signal and idler of a home-built three-stage optical parametric amplifier (OPA) pumped by the 1 kHz regenerative amplifier. Tuning the type II phase matching in the optical parametric amplifier allowed for generation of pulses at the desired wavelength of 11.6 μm, which were made narrowband downstream with a set of three identical commercial bandpass filters (Spectrogon BP11660-813). These pulses were made circularly polarized with a home-built phase retarder described in Supplementary Section S4. A ZnSe lens (f = 50 mm) was used to focus the pump pulses down to spot size of 80 μm $1/e^2$ diameter and struck the sample at normal incidence.

The probe pulses were focused by an achromatic doublet lens (f = 125 mm) down to a spot size of 40 μm $1/e^2$ diameter and were reflected by the sample with a ~7°incidence angle. The pump-induced changes in polarization rotation were measured using a balanced photodetection scheme, which consists of a half-wave plate (HWP) on a motorized rotation stage, a Wollaston prism, and a home-built balanced photodiode. The balanced photodiode records the average and difference of the intensities of the two beams separated by the polarization beamsplitter, and the HWP is rotated such that the two channels have equal intensities in the absence of a pump or applied magnetic fields. Any intensity imbalance measured upon excitation or application of a magnetic field then indicates a change in polarization state induced by the sample. Here, the measured polarizations are much smaller than $10^{-2}$ rad, and the detector response can be considered approximately linear in the rotation. The rotation angle can then be calculated as:

$$\Delta\theta = \frac{1}{2}\frac{A - B}{A + B} \tag{S3.1}$$

where A and B are the intensities of light in the two channels.

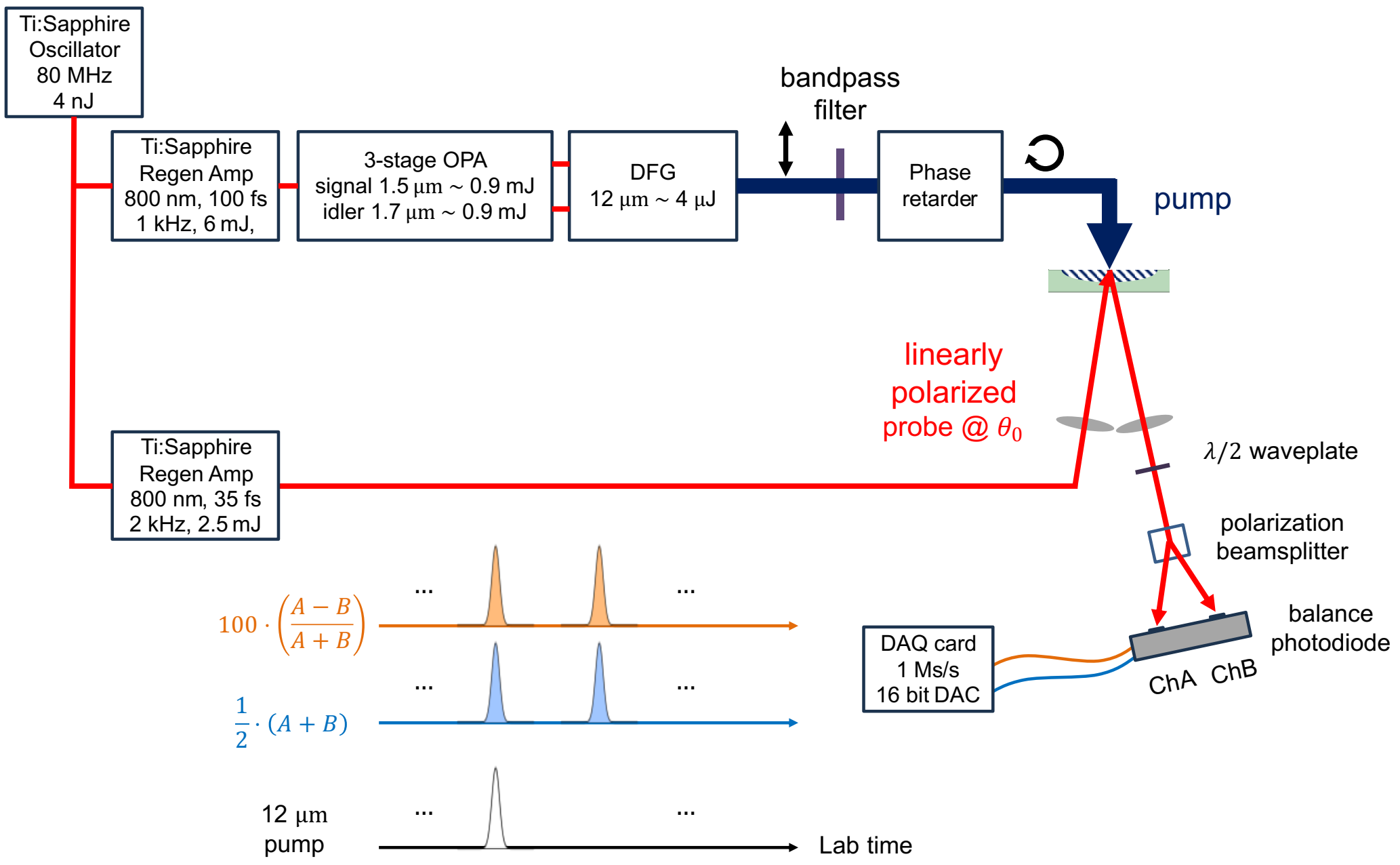

**Figure S3.1.** Experimental setup used for the transient polarization measurements.

To ensure that in the absence of the pump the polarization remained unchanged after traversing the sample and the focusing lenses we installed a piece of c-cut 5 mm thick sapphire (uniaxial axis along c axis, $n_o = 1.760$, $n_e = 1.752$ @ 800 nm) on a mirror mount with motorized actuators in the probe path before the sample. Tilting the crystal with the two actuators changes the angle between its optical axis and the probe propagation direction, and hence the refractive index seen by the extraordinary polarization, making the crystal a variable waveplate that is used to compensate residual static birefringence in the path.

A Data Acquisition (DAQ) card digitizes both the signals of the average and the difference channels from the balance photodiode. Each DAQ acquisition contains a train of pulses, which goes through a boxcar function that gates individual pulses, and the pulse-wise polarization rotation is calculated. The pump-induced changes in polarization state are measured by subtracting signals of neighboring probe pulses between pump on and pump off.

### S3.2 – Long-term Drift Compensations and Ultimate Noise Floor

To accumulate enough statistic and reach a polarization uncertainty of ~5 nrad per point as demonstrated in the data shown in Figure 4, the total acquisition time was ~110 hours (~9 hours/point). To prevent the effect of drifts several active stabilization loops were implemented in the data acquisition code. Both the temporal and spatial overlap were automatically re-optimized every ~3 hours. To do so, the ZnSe lens (f = 50 mm) focusing the pump on the sample was mounted on a two-axis motorized translation stage. The sample position was also computer controlled. This allowed periodically moving to a bare GaP area and scanning both pump-lens axes and finally the pump-probe delay to correct for possible pump pointing and temporal drifts. An example of the real-time evolution of the pump overlap parameters during a measurement is shown in Figure S3.2. Similarly, the retarder delay (Supplementary Section S4) was re-adjusted every ~3 hours by acquiring and fitting an interferogram, to minimize drift in the interference fringes and ensure constant circularity of the pump at the sample position.

At the probe pulse energy used in these experiments, each probe pulse delivered to the balanced photodetector ~$7 \times 10^8$ photons. A 9-hour acquisition at 2 kHz repetition rate is expected to enable a shot-noise limited polarization rotation uncertainty of ~ $2.35 \times 10^{-9}$ rad which translates in an uncertainty of $3.3 \times 10^{-9}$ rad on the pump-induced rotation changes calculated subtracting pump-on and pump-off data. The achieved noise level of $5 \times 10^{-9}$ rad is within factor of unity from photon shot noise.

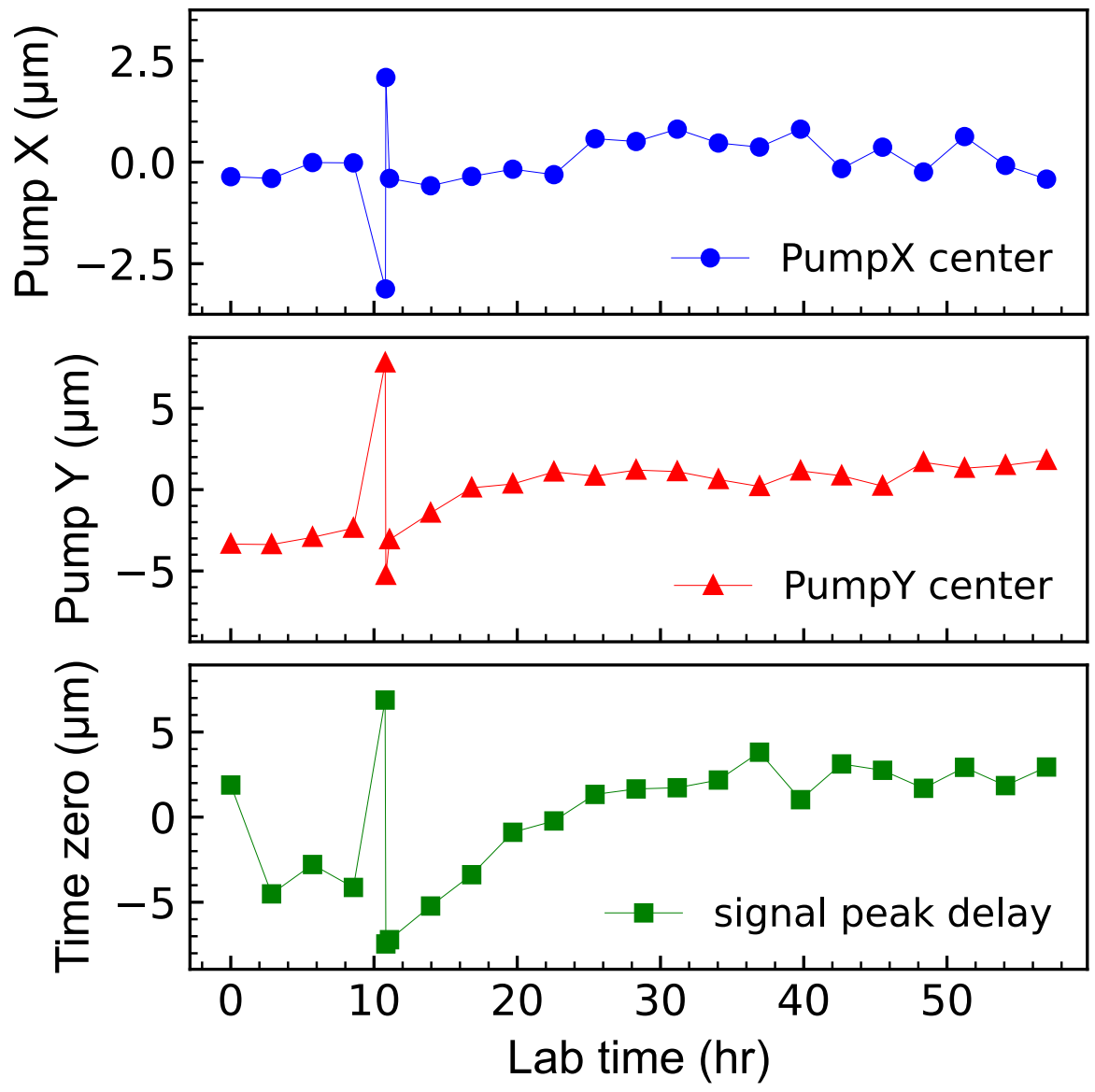

**Figure S3.2.** Real-time evolution of the pump overlap parameters over ~60 hours. The jump at the ~10th hour was caused by a manual re-optimization of the DFG signal-idler delay.

## S4. Generation of Circularly Polarized Pump Pulses

Circularly polarized light is generated by introducing phase retardation between two orthogonal linear polarizations using a home-built THz phase retarder. Two wire grid polarizers (Thorlabs WP25M-IRC) are used as polarization beamsplitters at an incidence angle of 15 degrees. The optical path length of one arm is tunable with a motorized delay stage (Physik Instrumente, NEXACT, N-565) with a precision better than 10 nm. Before the phase retarder, two additional wire-grid polarizers attenuate the pump intensity (together referred to as attenuator) and set the linear polarization angle entering the phase retarder. This allows one to control the relative intensity of the light circulating in the two time-delayed arms. The attenuator together with the retarder delay allows full control over the pump polarization state at the sample position.

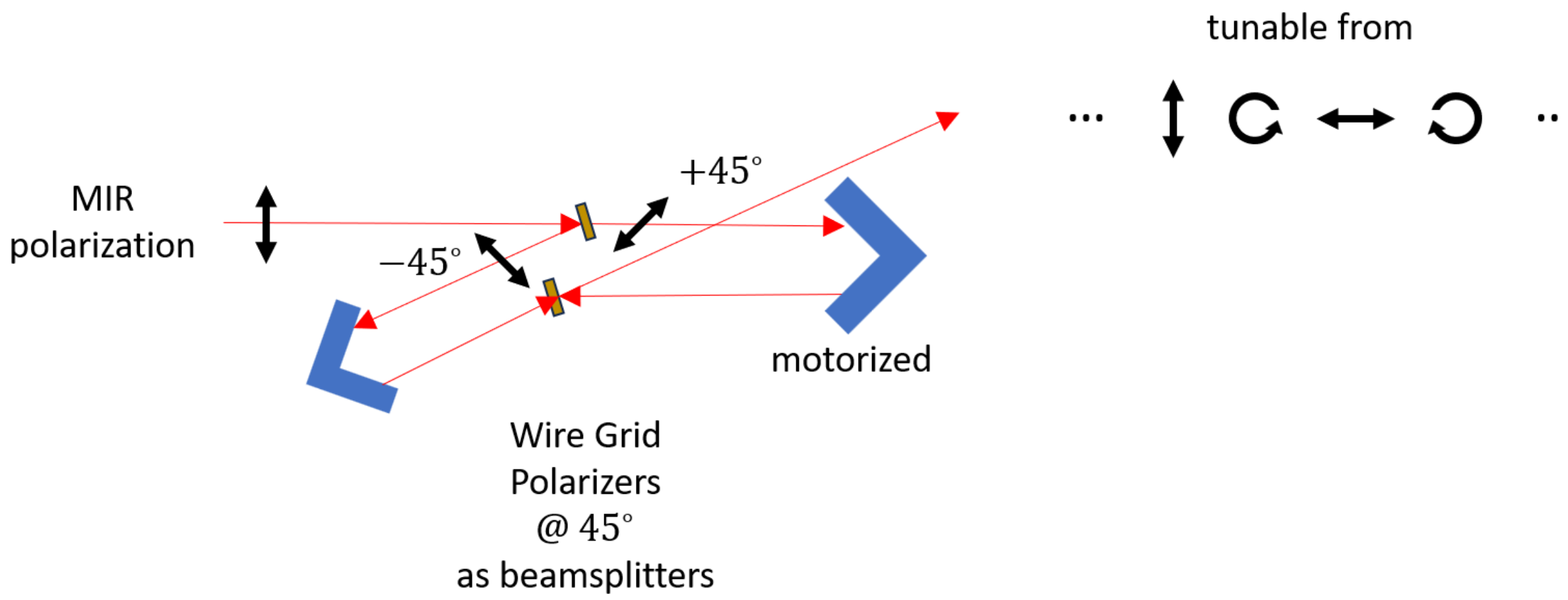

**Figure S4.1. Mid-infrared phase retarder schematic.** Incident polarization into the retarder is vertical, and the transmission axes of the two polarizers (Thorlabs WP25M-IRC) are oriented at 45° to the vertical.

The phase retarder can also be used to measure the pump spectrum. A parabolic mirror on a motorized flip mount intercepts the pump beam and focuses it on a wire grid polarizer transmitting vertical polarization. The transmitted beam intensity is recorded with a pyroelectric detector signal while scanning the retarder delay, yielding an interferogram from which the spectrum of the pump is obtained via Fourier transform.

The procedure for searching for a circular polarization configuration is as follows:

- Acquire a retarder delay ( $x$ ) scan, fit the interference fringes with $I(x) = Ae^{-\frac{(x-x_0)^2}{2\sigma^2}}\sin(kx-\phi_0)+I_0$ to determine the retarder delay at which temporal overlap is achieved.
- Place another wire grid polarizer (denoted as "analyzer") at the sample position where the pump light is focused, and measure the transmission intensity with a pyroelectric detector as a function of different analyzer angles.
- Perform a finer scan of the retarder delay around the stage positions that results in an interference intensity around half maximum, and fine tune the rotation of the second polarizer in the attenuator to optimize for the ellipticity of the pump polarization.

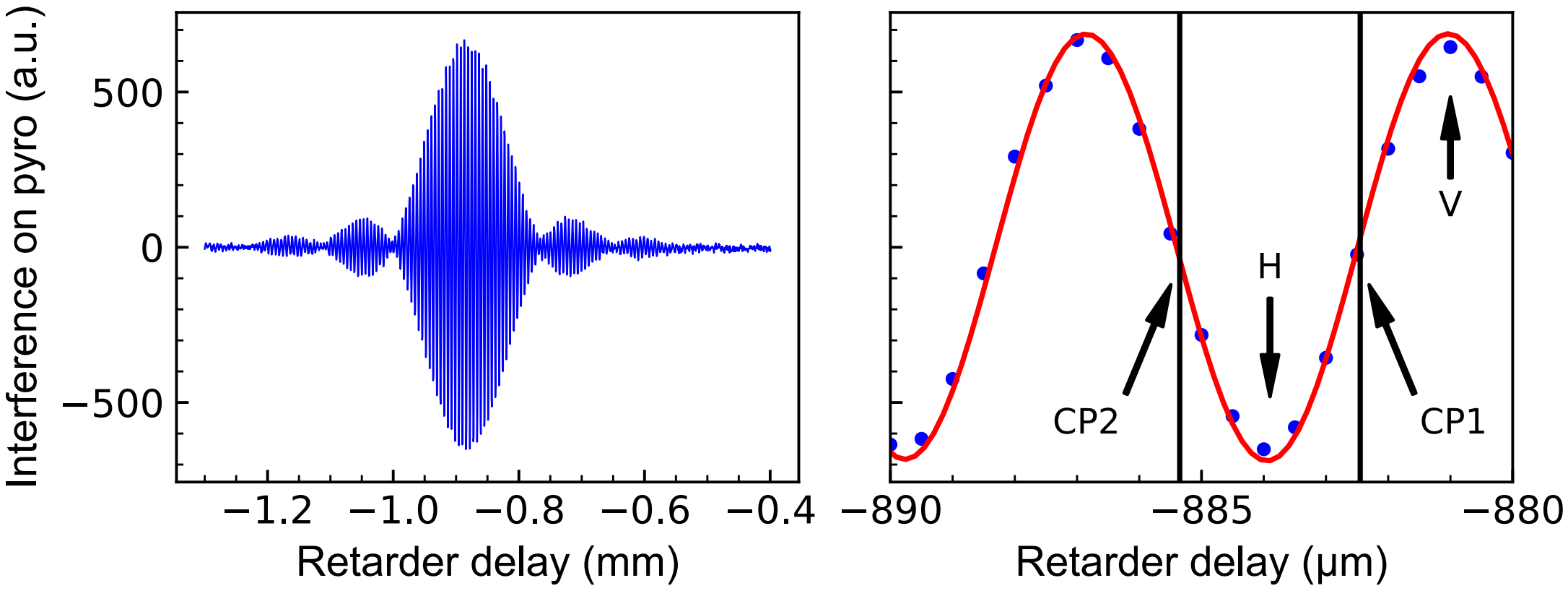

**Figure S4.2. Pump pulse interferogram and retarder delay choices for circular polarization.** (left) Complete interferogram of the pump pulse. (right) Zoomed in view of the interferogram near time overlap. The red curve is a local fit to the interferogram. Maximal constructive and destructive interference correspond to vertical (V) and horizontal (H) polarization exiting the retarder; the first and second retarder delay for circular polarization are found to the left of the delay for vertical polarization. These are denoted CP1 and CP2 and are calibrated to be $\sigma^+$ and $\sigma^-$ viewed along the crystal positive c-axis direction and accounting for the effect of the mirrors.

The maximal constructive interference happens when the two arms in the retarder that carries $\pm 45°$ polarization are of equal length. For consistency, the circular polarization configurations are always referenced to the zero value of the retarder delay. The first circular polarization configuration to the negative side of the vertical configuration is denoted as “CP1” (typically $\sim$ 22% of the interference period), the second is “CP2” (typically $\sim$ 71% of the interference period). The slight deviations from quarter and three-quarter interference period are introduced to compensate for the polarization changes introduced by metallic mirrors guiding light from the phase retarder to the sample. Figure S4.3 shows the transmission of the pump pulse through a polarizer placed at the sample position measured with a pyroelectric detector at different analyzer angles $\theta_0$. The data indicates an ellipticity of 0.95.

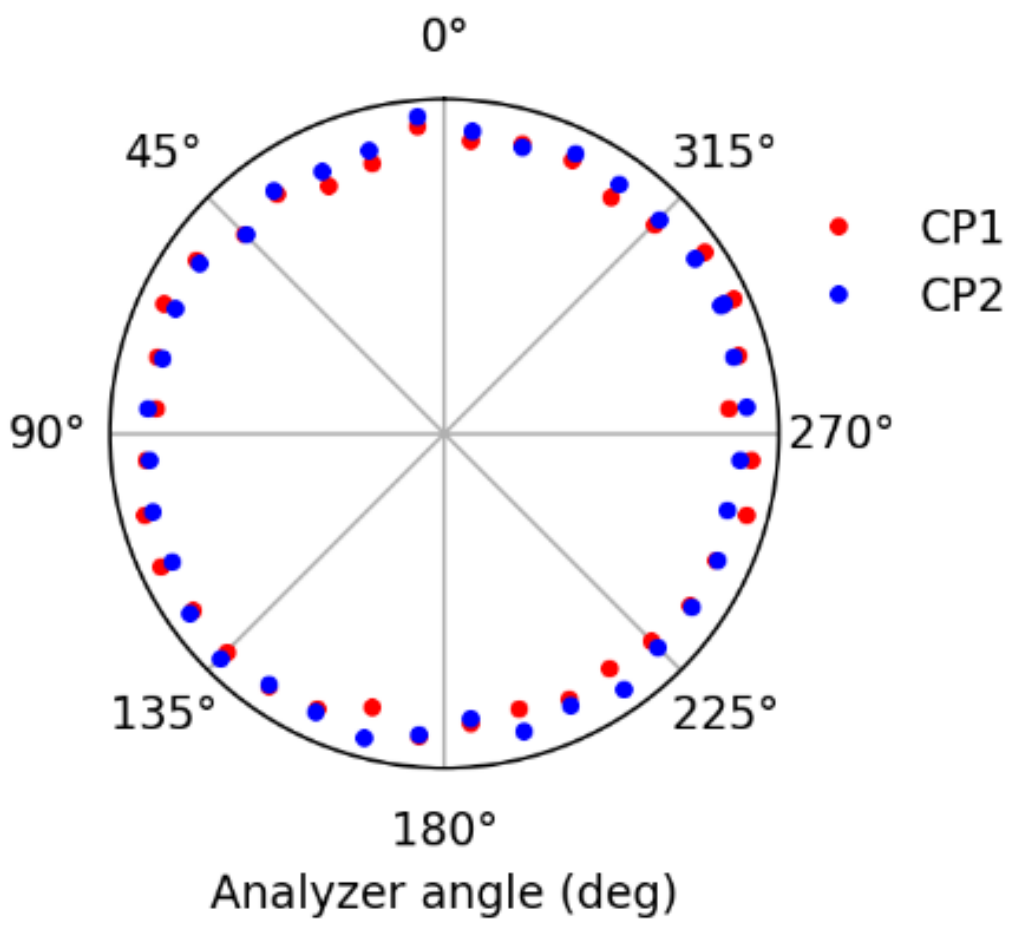

**Figure S4.3. Pump pulse circularity measurement.** The ellipticity is excellent for both circular polarizations CP1 and CP2 with opposite helicities.

## S5. Helicity dependent polarization rotation in 6H-SiC

Figure 3 of the main text reports the incident polarization-averaged half-difference of time-dependent polarization rotation scans acquired with opposite pump helicities. Figure S5.1 shows time dependent polarization rotation traces acquired for $\sigma^+$ and $\sigma^-$ pump helicities, obtained after averaging scans at incident probe polarization varying from 0 to 355° in 5° steps.

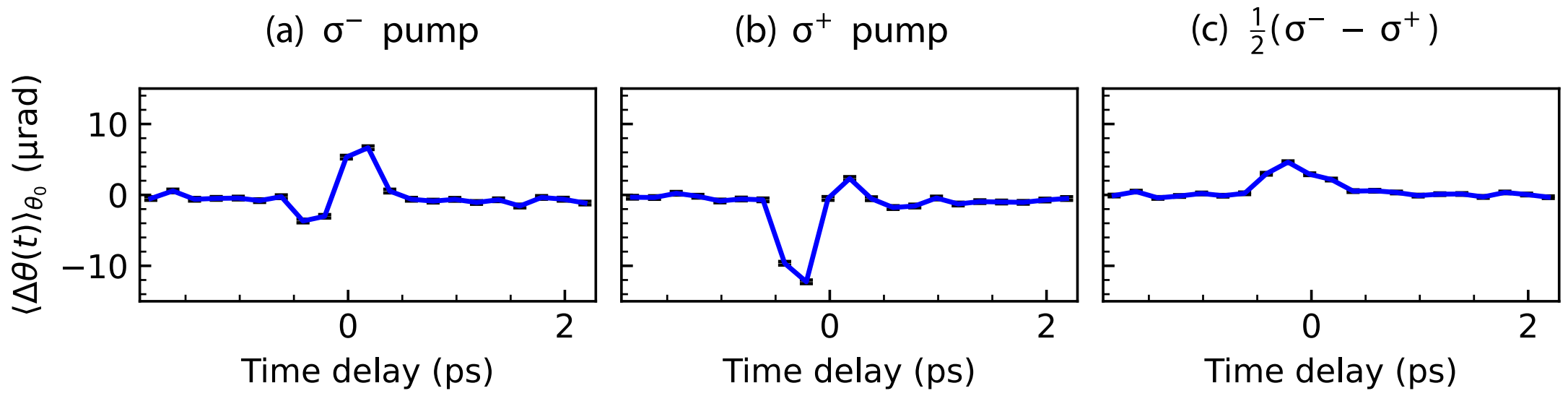

**Figure S5.1 Faraday rotation measurement in the photoexcited volume in SiC.** The three panels show the incident polarization-averaged polarization rotation measured in 6H-SiC for different pump helicities and the resulting half difference. These measurements were performed at room temperature.

## S6. SiC Device for Ultrafast Magnetometry

### S6.1 – Device Construction

To perform the measurements reported in Figure 4 and 5 of the main text a device multi-layer device was assembled on an insulating frame with a mm-sized window. To start with, a c-cut 6H-SiC was hand polished down to a thickness of 35 μm and was fixed with GE Varnish to the silicon frame. A wedged GaP (100) crystal was sourced from a commercial manufacturer (SurfaceNet GmbH) and was placed on top of the 6H-SiC crystal. The GaP crystal is 50 μm thick on the thin edge, with a 1.5 degrees wedge angle. The thick side of GaP is lifted from the silicon frame by a ~70 μm thick spacer, so that the front and the back surfaces of GaP are not parallel to the SiC surface. This ensured that only the light from the GaP interface closest to the SiC crystal is collected for polarization analysis.

### S6.2 – Choice of the 6H-SiC thickness

The choice of the 6H-SiC thickness resulted from a compromise between placing the detector as close as possible to the photo-excited volume and ensuring that the pump was fully shielded and did not reach the detector. Two different thicknesses were tested: ~15 μm and ~35 μm. To ensure that no signal would be detected due to pump-induced birefringence in the GaP crystal, we have acquired polarization rotation time scans with linear pump polarization and made sure that no signal was detected. While the stack built with the ~15 μm thick crystal showed a significant pump-induced birefringence, the ~35 μm crystal did not show any down to the detection limit shown in Figure 4 and was chosen as best compromise.

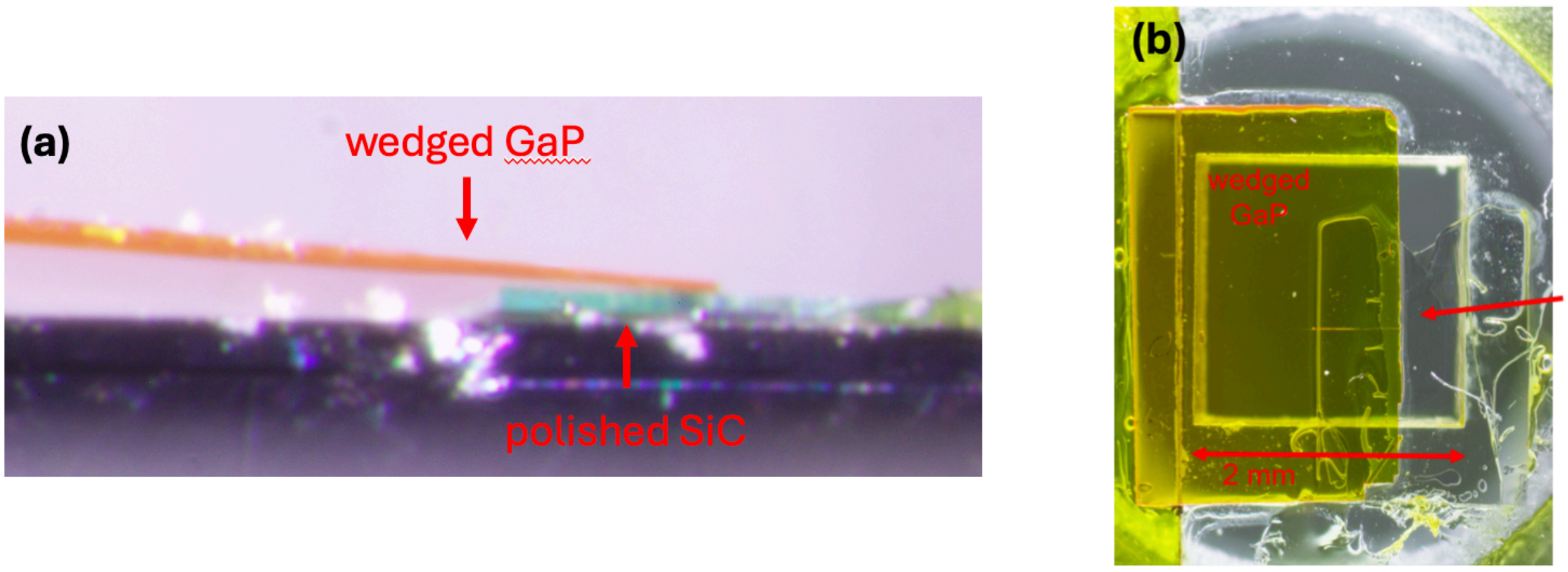

**Figure S6.1. GaP/SiC device for ultrafast magnetometry. (a)** Side view micrograph of the GaP detector (orange) stacked on the 6H-SiC sample (light blue). **(b)** Top view micrograph of the sample detector assembly.

## S7. Ultrafast Magnetometry Technique

Ultrafast optical magnetometry relies on the Faraday effect, which directly relates the magnetic field applied to a material to the polarization rotation of a linearly polarized beam traversing the medium. This relation is usually reported as:

$$\theta = V \cdot \int_0^L B(z)dz \qquad \text{(S7.1)}$$

where θ represents the rotation of the polarization of the input beam, B(z) is the magnitude of the magnetic field along the light propagation direction inside the medium, L is the thickness of the medium, and V is the Verdet constant. This expression highlights the fact that the magnetic field is averaged across the thickness of the material traversed. Therefore, if B varies strongly over distances of order ~L, its spatial dependence is blurred in the accumulated polarization rotation.

In the past, the Faraday effect in ferromagnetic crystals and thin films has been used to image the magnetic properties of superconductors at equilibrium[4,5]. These types of detectors (such as $Bi{:}R_3Fe_5O_{12}$, EuS, and EuSe) offer very high sensitivity ($V \sim 10^5$ $rad \cdot T^{-1} \cdot m^{-1}$) but have limited time resolution, down to 100 ps at best[6], due to the presence of low-lying magnetic excitations (e.g., ferromagnetic resonance) at sub-THz frequencies. On the other hand, diamagnetic II-VI and III-V semiconductors such as ZnSe, ZnTe, and GaP have a magneto-optic response featuring Verdet constants that are two to three orders of magnitude smaller than those observed in ferromagnetic materials but have the advantage of not being magnetically ordered and offer significantly better time resolution[6–8].

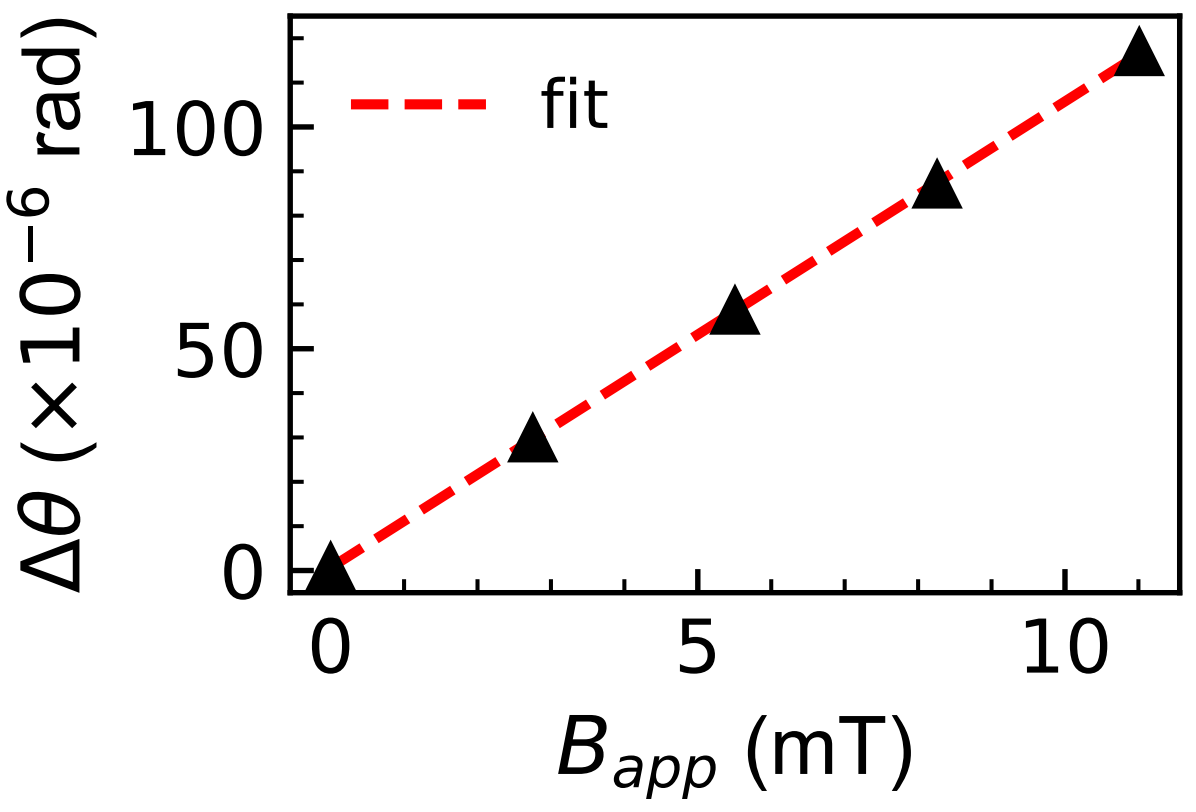

**Figure S7.1.** Sensitivity calibration measurement of the GaP (100) detectors used for the measurements presented in this work.

The measurements shown throughout the manuscript were performed using GaP detectors prepared as detailed in Supplementary Section S2. The sensitivity of these detectors was calibrated using the same polarization analysis setup used for the measurements. This was done recording the field-induced polarization rotation applying known magnetic fields with an in-situ Helmholtz coil pair which was independently calibrated using a Lakeshore 425 gaussmeter. The results of this calibration measurement, performed at room temperature is reported in figure S7.1 and accounting for the detector thickness yielded a Verdet constant of $\sim 120\ \mathrm{rad \cdot T^{-1} \cdot m^{-1}}$, in agreement with reported literature values. Note that for the specific (100) cut the Pockels effect is zero in GaP for electric fields in the plane of the detector crystal.

Figure S7.2 showcases the capabilities of GaP (100) as an ultrafast magnetic field detector. Quenching a $YBCO_7$ superconducting disc with a 400 nm laser pulse generates a magnetic field step with a picosecond risetime. The transient magnetic field is picked up by the GaP detector as shown in Figure S7.2b demonstrating a time resolution of ~1 ps, limited by the propagation time of the probe through the GaP crystal.

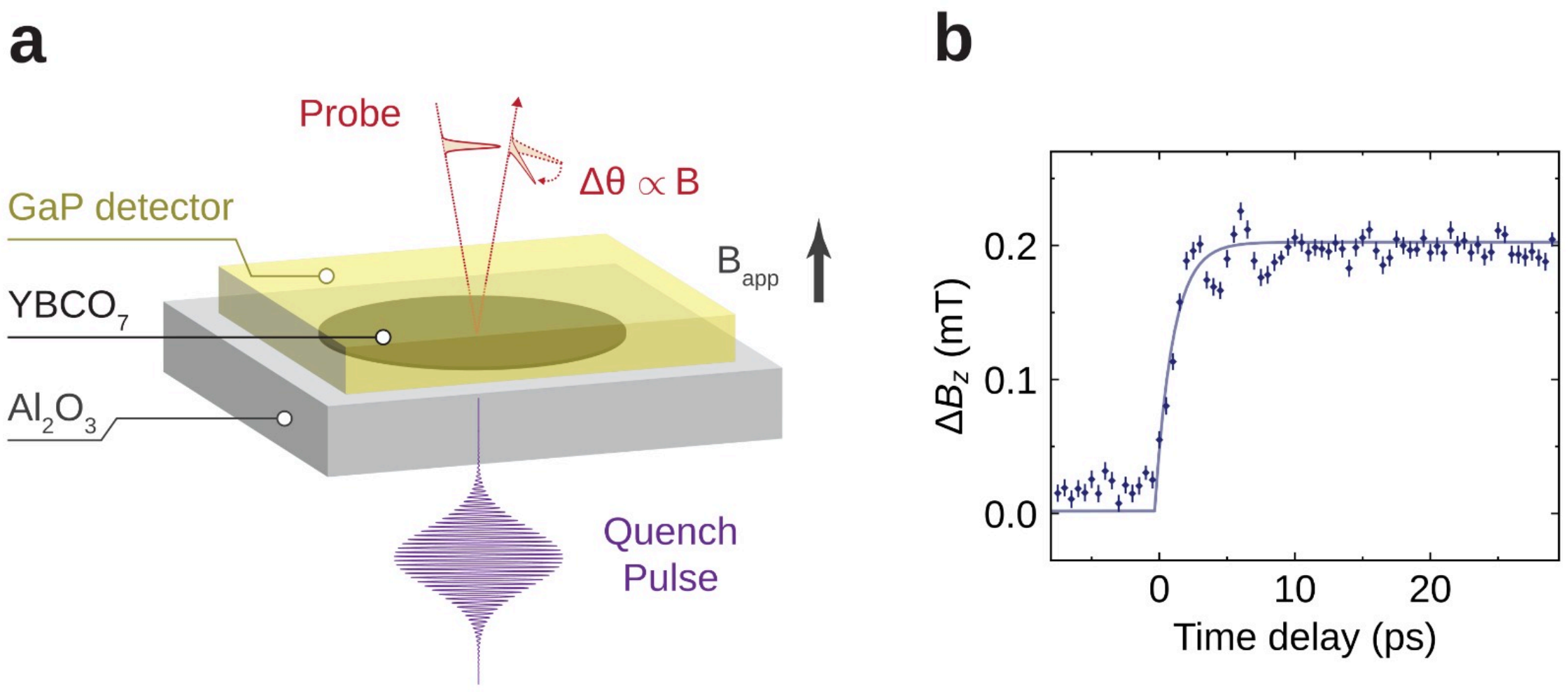

**Figure S7.2. Fast magnetic-field transient measured using GaP as a detector. (a)** Schematic of the experiment. Superconductivity in a YBCO thin disc on a $Al_2O_3$ substrate is disrupted using a 400 nm laser pulse and the magnetic field surrounding the disc is probed using the Faraday effect in GaP. **(b)** Time dependent magnetic field transient measured using the GaP (100) detector. The data is reproduced from Ref. [6].

## S8. Finite Element Simulation of the Magnetic Stray Field

To extract the stray magnetic field that would be picked up by the GaP magneto-optic detector if the photo-excited volume assumed a magnetization we used time-dependent finite element calculations solving Maxwell's equation using COMSOL Multiphysics. Given the radial symmetry of the problem, the simulations were performed in cylindrical coordinates. Figure S8.1a shows the geometry of the simulation. A 60 μm domain modelling the GaP crystal was defined and placed 5 μm away from the 35 μm domain modelling SiC. Inside it, a thinner domain models the photo-excited area. The effect of the phonon excitation is modelled as a time-dependent transient magnetization following the time profile shown in the top panel of Figure S8.2, reproducing closely the time-dependence of the polarization rotation measured inside the material. The spatial profile of the magnetization was assumed Gaussian, with a width equal to that of the pump beam diameter and homogeneous in the depth, with a thickness equal to the effective pump penetration depth. Figure S8.1(b-e) show color plots of the stray magnetic field generated around the photo-excited are and propagating away from it. All of the different materials included in the simulation are modelled using the appropriate dielectric functions.

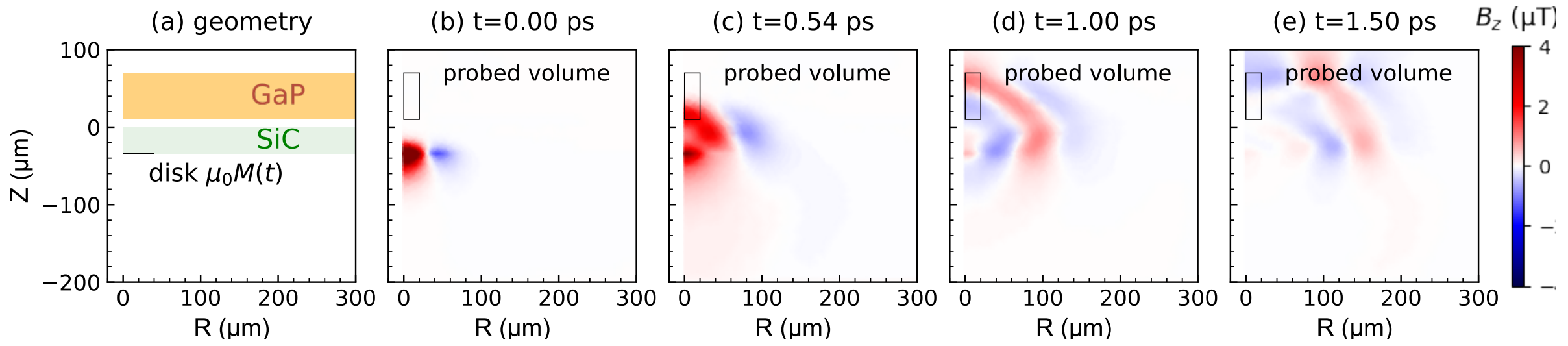

**Figure S8.1. Finite element simulation of the stray magnetic field. (a)** Geometry of the simulation. **(b-e)** Time evolution of the stray magnetic field around the photoexcited volume at different times measured from the peak of the magnetization response.

The probed volume (indicated in Figure S8.1(b-e)) is assumed to be equal to the thickness of the GaP detector and having a width of 40 μm, equal to the probe beam diameter. Averaging of the calculated magnetic field throughout this volume yields the calculated response measured by the GaP magneto-optic detector (Figure S8.2, lower panel). As discussed in the main text these simulations are repeated for different values of the peak magnetization in the photo-excited volume and a calibration curve relating the measured peak stray magnetic field to the peak magnetization acquired by the photoexcited volume (Figure 5b, main text).

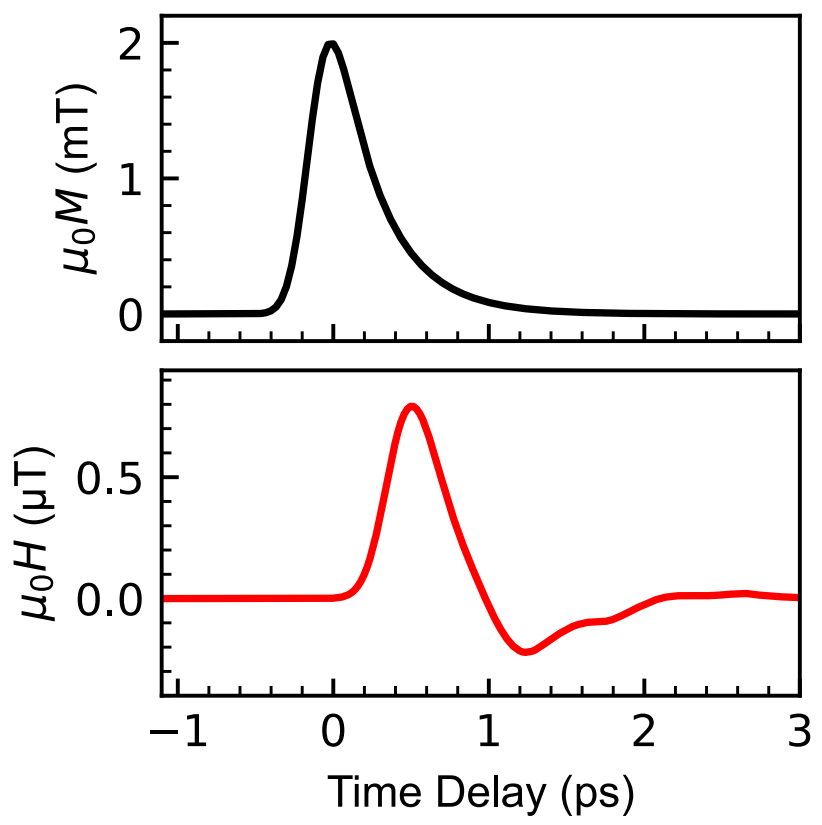

**Figure S8.2. (upper panel)** Assumed time-dependent magnetization profile. **(lower panel)** Calculated response in the GaP detector after averaging through the thickness of the detector. For a transient magnetization peaking at 2 mT, the stray magnetic field peaks at around 800 nT.

In addition, to benchmark the sensitivity of these calibration curves on the assumed pump penetration depth, we ran simulations for the following scenarios:

- $d = 0.25\ \mu m$, corresponding to the intensity penetration depth of the pump obtained as $\lambda/4\pi\ Im[\tilde{n}(\lambda)]$ where $\tilde{n}(\lambda)$ is the complex refractive index of 6H-SiC at the pump frequency.
- $d = 0.5\ \mu m$, corresponding to the electric field penetration depth of the pump and compatible with the saturating fluence dependence showed in section S9.
- $d = 0.75\ \mu m$, corresponding to the region where the pump energy is attenuated by $\sim$95%.

As can be observed from Figure S8.3 changing the assumption of the penetration depth between these values rescales the upper bound given for the phonon generated magnetization by no more than a factor of two relative to the assumption used in the main text of $d = 0.5\ \mu m$, leaving the conclusions of our manuscript unchanged.

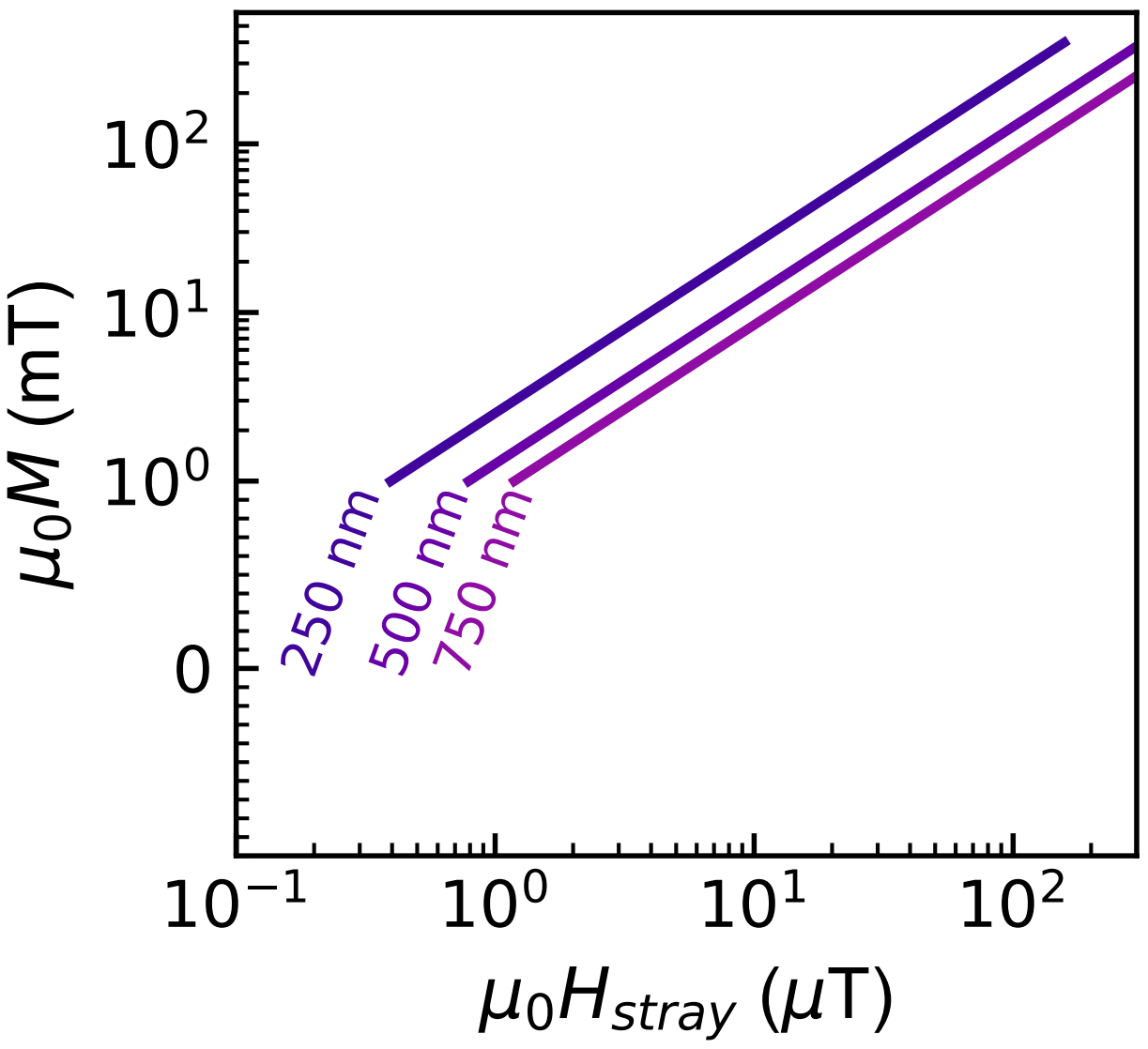

**Figure S8.3.** Calibration curves relating the stray field measured in the GaP detector to the expected peak value assumed by the magnetization in the photo-excited volume. The choice of the three different values is discussed in the text.

## S9. Fluence Depencence of Polarization Rotation in SiC

In Figure S9.1 we report the dependence of the measured polarization rotation on the excitation fluence. These data are acquired at the peak to the response, at room temperature, subtracting data acquired with opposite helicities. The peak polarization rotation shows a sublinear scaling with fluence. A possible reason for the observed sublinear scaling is a saturation with fluence of the driven phonon mode coordinate $Q$.

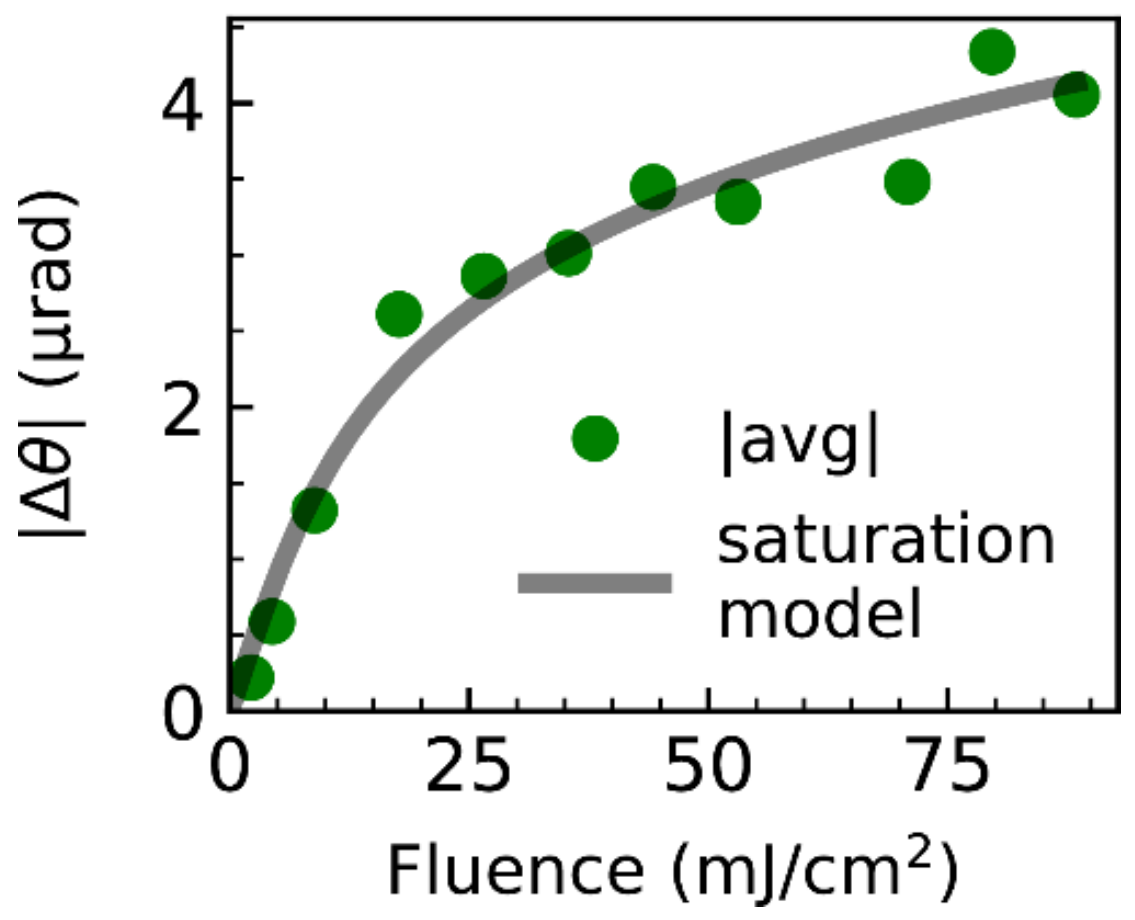

**Figure S9.1. Fluence dependence of the peak polarization rotation in SiC.** Green dots represent the peak of the incident-polarization-averaged polarization rotation measured in the SiC photoexcited volume. The gray line is a fit with a saturation model described in the text. This data was acquired at room temperature.

The observed behavior can be modelled assuming that the polarization rotation follows a saturation law:

$$|\Delta\theta| = \Delta\theta_{max} \frac{F}{F + F_{sat}}.$$

Performing a fit using Eq. XX, yields $F_{sat} = 6.17$ mJ/cm$^2$.

## References (Supplementary Material)